%% file: paper_v2.tex
\PassOptionsToPackage{unicode}{hyperref}
\PassOptionsToPackage{hyphens}{url}
\PassOptionsToPackage{dvipsnames,svgnames,x11names}{xcolor}
\documentclass[
  12pt,
  doublespace]{article}
\usepackage{xcolor}
\usepackage[margin=1in]{geometry}
\usepackage{amsmath,amssymb}
\usepackage{iftex}
\ifPDFTeX
  \usepackage[T1]{fontenc}
  \usepackage[utf8]{inputenc}
  \usepackage{textcomp} 
\else 
  \usepackage{unicode-math} 
  \defaultfontfeatures{Scale=MatchLowercase}
  \defaultfontfeatures[\rmfamily]{Ligatures=TeX,Scale=1}
\fi
\usepackage{lmodern}
\ifPDFTeX\else
\fi
\IfFileExists{upquote.sty}{\usepackage{upquote}}{}
\IfFileExists{microtype.sty}{
  \usepackage[]{microtype}
  \UseMicrotypeSet[protrusion]{basicmath} 
}{}
\makeatletter
\@ifundefined{KOMAClassName}{
  \IfFileExists{parskip.sty}{%
    \usepackage{parskip}
  }{
    \setlength{\parindent}{0pt}
    \setlength{\parskip}{6pt plus 2pt minus 1pt}}
}{
  \KOMAoptions{parskip=half}}
\makeatother
\makeatletter
\ifx\paragraph\undefined\else
  \let\oldparagraph\paragraph
  \renewcommand{\paragraph}{
    \@ifstar
      \xxxParagraphStar
      \xxxParagraphNoStar
  }
  \newcommand{\xxxParagraphStar}[1]{\oldparagraph*{#1}\mbox{}}
  \newcommand{\xxxParagraphNoStar}[1]{\oldparagraph{#1}\mbox{}}
\fi
\ifx\subparagraph\undefined\else
  \let\oldsubparagraph\subparagraph
  \renewcommand{\subparagraph}{
    \@ifstar
      \xxxSubParagraphStar
      \xxxSubParagraphNoStar
  }
  \newcommand{\xxxSubParagraphStar}[1]{\oldsubparagraph*{#1}\mbox{}}
  \newcommand{\xxxSubParagraphNoStar}[1]{\oldsubparagraph{#1}\mbox{}}
\fi
\makeatother

\usepackage{longtable,booktabs,array}
\usepackage{calc} 
\usepackage{etoolbox}
\makeatletter
\patchcmd\longtable{\par}{\if@noskipsec\mbox{}\fi\par}{}{}
\makeatother
\IfFileExists{footnotehyper.sty}{\usepackage{footnotehyper}}{\usepackage{footnote}}
\makesavenoteenv{longtable}
\usepackage{graphicx}
\makeatletter
\newsavebox\pandoc@box
\newcommand*\pandocbounded[1]{
  \sbox\pandoc@box{#1}%
  \Gscale@div\@tempa{\textheight}{\dimexpr\ht\pandoc@box+\dp\pandoc@box\relax}%
  \Gscale@div\@tempb{\linewidth}{\wd\pandoc@box}%
  \ifdim\@tempb\p@<\@tempa\p@\let\@tempa\@tempb\fi
  \ifdim\@tempa\p@<\p@\scalebox{\@tempa}{\usebox\pandoc@box}%
  \else\usebox{\pandoc@box}%
  \fi%
}
\def\fps@figure{htbp}
\makeatother

\usepackage[]{natbib}
\usepackage{amsmath}
\usepackage{amssymb}
\usepackage{booktabs}
\usepackage{tabularray}
\UseTblrLibrary{booktabs}
\usepackage{longtable}
\usepackage{graphicx}
\usepackage{csquotes}
\usepackage{setspace}
\let\oldabstract\abstract
\let\endoldabstract\endabstract
\renewenvironment{abstract}{\oldabstract\singlespacing}{\endoldabstract}
\usepackage{fancyhdr}
\usepackage{titlesec}
\titleformat{\section}{\normalfont\bfseries}{\thesection.}{1em}{}
\titleformat{\subsection}{\normalfont\itshape}{\thesubsection.}{1em}{}
\titleformat{\subsubsection}{\normalfont\itshape}{\thesubsubsection.}{1em}{}
\usepackage{caption}
\usepackage{hyperref}
\hypersetup{
  colorlinks=true,
  linkcolor=blue,
  citecolor=blue,
  urlcolor=blue
}
\usepackage{lscape}
\makeatletter
\@ifpackageloaded{caption}{}{\usepackage{caption}}
\AtBeginDocument{%
\ifdefined\contentsname
  \renewcommand*\contentsname{Table of contents}
\else
  \newcommand\contentsname{Table of contents}
\fi
\ifdefined\listfigurename
  \renewcommand*\listfigurename{List of Figures}
\else
  \newcommand\listfigurename{List of Figures}
\fi
\ifdefined\listtablename
  \renewcommand*\listtablename{List of Tables}
\else
  \newcommand\listtablename{List of Tables}
\fi
\ifdefined\figurename
  \renewcommand*\figurename{Figure}
\else
  \newcommand\figurename{Figure}
\fi
\ifdefined\tablename
  \renewcommand*\tablename{Table}
\else
  \newcommand\tablename{Table}
\fi
}
\@ifpackageloaded{float}{}{\usepackage{float}}
\floatstyle{ruled}
\@ifundefined{c@chapter}{\newfloat{codelisting}{h}{lop}}{\newfloat{codelisting}{h}{lop}[chapter]}
\floatname{codelisting}{Listing}

\makeatother
\makeatletter
\@ifpackageloaded{caption}{}{\usepackage{caption}}
\@ifpackageloaded{subcaption}{}{\usepackage{subcaption}}
\makeatother
\usepackage{bookmark}
\IfFileExists{xurl.sty}{\usepackage{xurl}}{} 
\hypersetup{
  pdftitle={Learning about Markets from Models: Evidence from an
AI Hiring Platform},
  pdfauthor={Hossein Alidaee; Tarun Khanna; Rem Koning},
  pdfkeywords={Labor market matching, Hiring platforms, Psychometric
assessment},
  colorlinks=true,
  linkcolor={blue},
  filecolor={Maroon},
  citecolor={Blue},
  urlcolor={Blue},
  pdfcreator={LaTeX via pandoc}}

\title{Learning about Markets from Models: \\ Evidence from an AI Hiring Platform\thanks{We gratefully acknowledge financial support from the HBS AI Institute and computing cluster support by the Harvard FAS Division of Science Research Computing Group. We thank Jorge Tamayo, Chiara Farronato, and participants at the University of Washington economics seminar and HBS Digital Research Day for helpful comments and discussions. Both the underlying data analysis and manuscript benefited from use of several AI tools, particularly OpenAI Codex, Claude Code, and Superwhisper.

\textsuperscript{1} Harvard Business School 

\textsuperscript{*} Correspondence: \href{mailto:halidaee@hbs.edu}{halidaee@hbs.edu} 
}}
\author{Hossein Alidaee\textsuperscript{1,*} \and Tarun
Khanna\textsuperscript{1} \and Rembrand Koning\textsuperscript{1}}
\date{\today}
\begin{document}
\setcounter{footnote}{1} 
\maketitle
\begin{abstract}

Machine learning models are increasingly used to not only inform, but make decisions. Sometimes platforms add value by pooling data across contexts to improve models' decision making---but when the optimal decision differs by context, pooling may not increase accuracy. We argue that a platform storing starkly distinct models across contexts nevertheless captures a \textit{model comparison externality}. Together, these models reflect how decision makers respond differently to the same data in the current equilibrium. Policymakers can use this differentiation to prioritize interventions; firms can learn how they are already differentiated from competitors to focus future investments. Our evidence comes from an AI hiring platform used by large employers of low-skill workers. The platform uses a standardized pyschometric assessment to create firm-specific models estimating each application's retention likelihood at that firm. Each firm's model is highly predictive for itself but no better than chance at other firms. We use these models to create a firm-application matrix: each firm scores each application, not only the ones it receives. An application's improved retention likelihood if submitted to a different firm is highly heterogeneous based on traits. Further, these gains are due to two distinct mechanisms. Finally, firms value traits differently even in this low skill market: they disagree about whether important traits improve or worsen retention at the margin. These insights are only possible when a platform can access both the full application pool and each firm-specific model. We explain the value of these insights to firms, policymakers, and platforms. 
\end{abstract}

\section{Introduction}\label{introduction}

The 21st century has witnessed the proliferation of data-informed decision making. Consumers receive algorithmic movie recommendations \citep{FlederHosanagar2009,CheHorner2018}. A/B tests inform firms’ product decisions \citep{azevedo2020,koning2022}. Machine learning models screen job applications \citep{hoffmanKahnLi2018,LiRaymondBergman2025}.  This data, and the models converting them into insights, may be managed locally or through a central platform. Pooling data through a central platform has clear benefits when a single model works well across contexts, as it improves statistical precision, creating a positive data sharing externality \citep{BergemannBonattiGan2022,HagiuWright2023}. For example, a firm bidding on digital ads benefits from a central platform that can aggregate user data across many clients to improve statistical precision of its estimates \citep{WernerfeltEtAl2025}. Yet, when others’ data don’t improve the precision of our own decision, is an externality still present? 

We argue that even in highly heterogeneous contexts, where one client’s data is incredibly informative for themselves and has no predictive power for others, an externality still exists. Unlike previously documented data sharing externalities, this externality is not because sharing data improves any given user’s predictions. Instead, we introduce a \textit{model comparison externality}: when economic decisions are delegated to models \citep{shahidi20266,liang2026clones}, these models can inform how different decision makers respond to the same circumstances under the current equilibrium. Understanding the differences across models produces valuable insights for firms, platforms, and policymakers about how to allocate resources. These insights would not be available if models were maintained internally by each decision maker instead of on a centralized platform. 

Consider the case of employers A and B who have trained models on their own hiring data to score job applicants. An applicant who receives a low score from A’s model may receive a high score from B’s. Having done this comparison, a platform can use this information to better match applicants to opportunities. For policymakers, comparing models can reveal if job seekers tend to get low scores across both firms—suggesting a need for training programs—or if scores depend on where applicants apply—suggesting a need for better job-search and placement assistance programs. There are benefits from pooling data and models even when doing so leaves each model's predictions unchanged.

We use data from a hiring platform that is uniquely well suited to illustrate this model comparison mechanism. The platform's clients are large formal sector employers in Mexico. The platform administers a common psychometric assesssment to new employees who went through the firm's existing hiring process, then records their tenure at the company after 90 days. Instead of judging applications based on raw scores \citep[e.g.][]{autorScarborough08}, these data are used to train a firm-specific machine learning model. This model then evaluates future applications, estimating the probability their tenure will exceed 90 days using only this psychometric data. Low skill service sector workers have well-documented high levels of attrition, which is costly to firms \citep{hoffmanKahnLi2018, kuhnYu2021}, making the estimates from these models especially valuable to hiring decisions. Since applicants all take the same psychometric assessment, we can use these models to also test our proposed externality. First, we can show that models are highly predictive but only at their intended firm. Subsequently, we can show that access to multiple models can generate policy insights about worker-firm complementarities, relevant to both firms and policymakers. 

Specifically, we use data from three of the platform's largest clients covering a pool of 172,242 job applications from 2020-2025. Using this data we present three sets of results. First, we show a model trained on a firm's own data is highly predictive at ranking applications: at each of the three firms in our sample, they rank a randomly selected worker who stays above one who leaves at least 85\% of the time. However, we find that a model from Firm A is essentially useless at predicting turnover for workers at Firm B. While data is very valuable within a given firm, in this case it is much less useful in other contexts, pointing to limited data sharing externalities in this setting.

Despite these models having no predictive value across contexts, our second set of results illustrate that access to multiple models can nevertheless generate other benefits. In our setting, by comparing models across firms, platforms and policymakers can observe improvements in predicted tenure from redirecting applications to a more suitable firm. \citet{Jovanovic1979} first documented the challenges in estimating how a given worker's retention changes based on where they work. Our platform setting, unique because of the standardized data on workers' traits and retention outcomes---and because it includes emperical evidence that these are linked---helps overcome this challenge. We use our data to create an application-firm matrix, estimating the predicted retention of each application in our sample at each firm in our sample, and then  calculate each application's gain between its best score across firms versus where it was actually submitted. We find that an application's average score across the seven psychometric traits is heavily related to the gain they might expect from applying to a different firm. Applications scored in the top quintile see a predicted retention probablilty increase of 26.9 percentage points; applications in the bottom quintile only see a 16.4 percentage point gain from application redirection. These results suggest that matching frictions in our labor market may heterogeneously impact workers based on their psychometric traits. 

These differences originate from two possible ways an application's predicted retention would improve if routed to its best-fit firm. First, the best-fit firm could place higher value on that application's unique combination of traits, implying genuine worker-firm complementarities. Second, that best-fit firm may have a higher baseline retention across all profiles than the firm to which they actually applied. These types of gains arise from a mechanism known as sorting without complementarities \citep{AndersonSmith2024,ChadeEeckhoutSmith2017}. To account for the latter, we adapt the share-matching inversion used in discrete-choice demand estimation \citep{train2009discrete, berry1995automobile} to our binary machine learning estimators. We recalibrate each model to have the same average predicted retention while preserving differences in log-odds across applications. Recalibration affects applications differently across quintiles: the top quintile sees a 7 percentage point drop in predicted gains from redirection, while the bottom quintile's gains only decrease by 2.2 percentage points. Together, these results imply that workers are the top of the distribution are the ones most impacted by both mechanisms: worker-firm complementarities and sorting without complementarities. While this mechanism has been recently studied in the theoretical literature, we are, to our knowledge, the first to document its importance in a labor market. These findings suggest that reallocation policies may have different relative impacts across applications, depending on which mechanism a policy emphasizes. 

Finally, to explore potential explanations of these gains, we use two different methodologies for parameterizing the high-dimensional, non-linear relationships within these models. First, Grouped Shapley values \citep{JullumRedelmeierAas2021, LundbergLee2017} calculate, for each model, each trait's average absolute contribution to predicted retention, in percentage points, by comparing predictions using an application's trait data with an average over predictions using other applications' data for that trait. Firms' show modest disagreements in aggregate rankings of traits, but broadly agree in certain comparisons: collaboration is one of the most important traits while diligence is among the least. Second, to estimate--and conduct inference on---the marginal effect of each trait on predicted retention, we implement \citet{ChernozhukovNeweySingh2022}'s AutoDML methodology. Among the subset of traits with statistically significant estimates, traits that increase predicted retention at one firm decrease it at another. Moroever, this disagreement is sensitive to an application's score on that trait. For example, increasing raw emotional control scores among the middle quintiles has a statistically significant positive predicted effect at two firms and negative at the third. Yet, at the bottom and top quintiles, the effect is insignificant across all three firms' models. These results are consistent with broader empirical literature \citep[e.g.,][]{BensonLiShue2019, DemingKahn2018,BohmEsmkhaniGallipoli2025} validating \citet{Lazear2009}'s theory of firm-specific human capital, where firms are horizontally differentiated in which traits they value. Unlike prior work, our models do not impose linear or monotonic relationships between assessment and retention, and find that firms' best fitting mappings do not have these linear monotonic properties.

Our proposed mechanism, and our empirical findings, have important implications for firms, policymakers, and platforms. For firms, the existing literature documents that standardized assessments can improve job application screening \citep{autorScarborough08, Weidmann2026} and that managers overriding judgements from these assessments do so at their own detriment \citep{hoffmanKahnLi2018}. We document the potential for improving screening by allowing for nonlinear relationships between the traits measured in these assessments and desired outcomes. Further, we demonstrate that these nonlinear relationships look different across firms. While others have documented heterogeneity in what firms screen for on the job market \citep[e.g.][]{BensonLiShue2019, DemingKahn2018, BohmEsmkhaniGallipoli2025}, we demonstrate this is true even for so called ``low-skilled'' positions. Thus, firms may benefit from more targeted recruitment strategies even for these positions. However, understanding which trait profile a firm values most uniquely requires a broader understanding of the labor market. This emphasizes the importance of our model externality mechanism, and firms' trade-off between developing in-house models versus using centralized platforms that can produce broader insights by comparing models. 

For policymakers, we provide two distinct categories of insights. First, our results contribute to a literature seeking to understand frictions in the low skill service labor market. In developing economies like Mexico's, these frictions are especially acute: exit rates are high, transitions to better jobs are rare, and credible signals of worker quality are scarce \citep{donovan2023, dubeLesterReich2016}. Creating policies to address these concerns requires understanding the mechanism behind these frictions. Whether the population has the trait distribution firms want, versus needing to improve those traits, affects whether policymakers should invest in platforms directing workers to their best fit employer \citep{carranzaGarlickOrkinRankin2022, bassi2022, beamanMagruder2012, wheeler2022} versus training programs  improving the quality of the local labor supply. Our results indicate substantial heterogeneity in the importance of firm fit across the talent distribution; this suggests policymakers' optimal intervention may need to be context specific \citep{Allcott2015, HotzImbensMortimer2005}. 

Our results also inform how policymakers should pursue platform investment and regulation. While small third-parties doing labor force skill assessment can effectively reduce the uncertainty in personnel decisions \citep[e.g.][]{carranzaGarlickOrkinRankin2022}, they provide firms with the raw scores of applications instead of locally adapted predictions. Yet, because the relationship between assessments and outcomes is both non-linear and highly context-specific, redirecting investments towards a centralized platform that can learn these relationships and deliver context-specific predictions may have greater impact on firm behavior \citep{Alidaee}. While prior work on labor platforms has highlighted their policy role in alleviating search frictions \citep{Horton2017Recommendations} and helping firms vertically differentiate the workforce \citep{KrasnokutskayaSongTang2020, Yoganarasimhan2013}, we believe ours is the first to illustrate how centralized platform can inform policymakers about the relative importance of vertical versus horizontal differentiation in the labor market. Furthermore, anti-trust efforts seeking to break up large platforms should consider the value of capturing model externalities, as they produce information valuable not only to the market's participants but to its policymakers as well. 

For platforms, we provide evidence for a novel mechanism about when pooling data is valuable. Past work emphasizes that data pooling creates externalities by improving local prediction rules from theoretical \citep{JonesTonetti2020, BergemannBonattiGan2022, GalpertiLevkunPerego2024} and empirical \citep{Yoganarasimhan2019} perspectives. Informed by a growing literature on the importance of heterogeneity for platform prediction and matching \citep{Yoganarasimhan2019,BondiRossiStevens2025, Hitsch2010}, we demonstrate that pooling data gives platforms insights beyond improving local predictions. By learning the differences in patterns across contexts, platforms can learn important insights impacting their value not only to clients, but to policymakers as well. 

Lastly, our paper illustrates how analyzing multiple models are an important methodological approach for scholars, managers, and policymakers. A growing body of research uses models, particularly LLMs, as synthetic individuals to study topics from auction design to behavioral economics to firm strategy \citep{Horton2023,jackson2025ai,tranchero2026}. However, models are not only useful as simulacra; they increasingly make economic decisions for firms and consumers \citep{shahidi20266,liang2026clones}. In our setting, each firm relies on its own supervised model to make hiring decisions; more broadly, consumers and firms now delegate tasks---ranging from purchasing goods to writing contracts---to AI agents. Thus, models are not only stand-ins for economic agents \citep{manning2026general}; they are actual participants in markets. Studying and comparing them offers a new way to directly generate field evidence on the decision rules that structural estimation has typically been used to infer.


The paper proceeds as follows. Section~\ref{sec-data} describes the platform, common assessment, and evaluation and production samples. Section~\ref{sec-local-prediction-portability} explains how we estimate firm-specific retention models and compare the rankings and retention probabilities they assign to the same applications. Section~\ref{sec-portability-evidence} shows whether, and by how much, the models disagree when scoring those applications. Section~\ref{sec-model-explainability} examines the sources of this disagreement: which assessment domains drive each model's predictions and how the same domain scores relate to retention across firms. Section~\ref{sec-conclusion} concludes.

\section{Data}\label{sec-data}

Our data are from several clients of a digital hiring platform in Mexico. Each client uses the platform to recruit workers for a variety of low-skill service positions, such as warehouse workers, truck drivers, or retail workers. From the platform's client pool, we selected firms that use the platform's own psychometric assessments, as opposed to a separate set of clients using firm-specific, proprietary assessments. This choice lets us score each application not only with the model of the firm to which it was submitted, but with the other firms' models as well.

The sample consists of three firms, each remaining confidential. Firm A is a multi-national consumer product goods firm, hiring for a variety of positions across its logistics chain from warehousing to last-mile delivery. Firm B is a multi-national coffeehouse chain, predominantly hiring for retail positions. Firm C is an international softline retail chain. All three firms have a significant interest in reducing their workforce's 90-day attrition rate, and explicitly hire this platform to train models to screen applicants with the goal of reducing this rate.

The platform data are organized at the application level. For each firm in our sample, we divide these applications into two datasets: (i) applications with observed assessment responses and attrition outcomes (\emph{evaluation data}), collected as part of the platform's onboarding process before the firm began using the assessment for screening, and (ii) applications screened through the platform after onboarding (\emph{production data}). For evaluation applications, we observe the assessment responses, the 90-day attrition outcome, and the firm in which that outcome was realized. For production applications, we observe the assessment responses and the firm receiving the application. The platform has limited additional information: gender, age, and education are available but enter neither our predictive models nor the platform's models. Applicants' work histories do not exist in any version of the data.

\subsection{Assessments}\label{sec-assessments}

The platform's assessment, administered via a WhatsApp chatbot, is divided into seven sections, each measuring a distinct behavioral trait, and each composed of six questions. These sections are: commitment to work, emotional control, rule adherence, ethics, collaboration, diligence, and safety at work. Each question in the assessment is a Likert scale question, with responses ranging from 1 (strongly disagree) to 5 (strongly agree). Although individual questions are proprietary, they contain standardized wording and are similar to other assessments estimating similar industrial psychometric traits. A sample question, similar to but not identical to an English translation of a question used in the platform's own assessment, is:

\begin{displayquote}
 ``\emph{I have an easy time following every instruction given to me.}''
\end{displayquote}

This item would belong to the assessment's rule adherence section. In addition to the responses to individual questions, we also observe the aggregated composite score for each section constructed by the platform. These aggregations are not tied to outcomes; they are simply normalized sums of responses to individual questions, given different questions have different ordinal rankings. For example, a higher Likert scale response to one question in rule adherence may imply high adherence while the phrasing of a different question in the same section may imply low adherence, thereby requiring adjustment.

This 42-question assessment was developed directly by the platform and is identical across all firms, and identical across production and evaluation data. Thus, every application record in our analysis lives in an identical feature space. This standardization is a requirement for our ability to pool data and estimate differences across models used by each firm.

\subsection{Production and Evaluation Data}\label{sec-production-and-evaluation-data}

Throughout the analysis, the unit of observation is an application. The evaluation sample is the exact application-level sample used to develop the prediction models. For all three firms, an evaluation application must have an observed binary 90-day attrition outcome, be associated with a worker recorded as hired and active rather than rejected, have no prior platform fitness score, and include observed responses to all 42 common assessment items and seven composite scores. The production sample retains applications with nonmissing responses to all 42 common items and seven composite scores and no observed attrition outcome. These restrictions yield 180,361 applications across the two samples.

The evaluation sample contains 8,119 applications. It is constructed before firms adopt the platform's screening tools. During onboarding, firms continue hiring without these screening tools, while assessed workers complete the common assessment and HR records their 90-day retention outcomes. For every evaluation application, we therefore observe the firm, responses to all 42 common assessment items and the seven composite scores, and whether the associated worker remained employed beyond 90 days. The samples were collected over different periods because the three firms joined the platform at different times.

The production sample contains 172,242 applications submitted after firms adopt the platform's screening tools, regardless of whether they resulted in a hire. For these applications, we observe the receiving firm, assessment responses, and demographics, but no retention outcome.\footnote{Hiring outcomes are also observed in this sample but not used.}

The two samples serve distinct purposes. We split evaluation applications into training and test folds to develop each firm's attrition model. We then use production applications as a common out-of-sample population on which to compare predicted retention across firm environments.

\subsection{Descriptive Statistics and the Cross-Firm Matrix}\label{descriptive-statistics-and-the-cross-firm-matrix}

Table \ref{tbl-table1} summarizes the scale and composition of the evaluation and production samples across the three firms.

\input{analysis/output/paper_evaluation/production/table1_restricted_samples.tex}

The evaluation sample contains 8,119 applications, while the production sample contains 172,242 applications. Firm C contributes the largest share of both samples. Among evaluation applications, the share with retention beyond 90 days is 35.8 percent at Firm A, 66.1 percent at Firm B, and 41.2 percent at Firm C. Table \ref{tbl-table1} also reports a limited set of demographic characteristics to describe the two samples. Although these characteristics vary across firms and between evaluation and production applications, they do not enter our predictive models; all model estimates use only responses to the common psychometric assessment and the composite scores constructed from those responses.

The two samples play complementary roles in the platform's multi-firm data architecture. Evaluation applications provide the firm-specific retention outcomes used to estimate each local prediction model, while production applications provide the common population on which those models are compared. Figure \ref{fig-application-firm-matrix} illustrates the resulting application-by-firm prediction matrix. Each row represents a production application, grouped by the firm to which it was submitted, and each column identifies the firm-specific model used to score it. Using its own data alone, each firm can populate only the diagonal block containing predictions from its model for its application pool. The platform can instead apply every firm-specific model to every production application, filling the off-diagonal blocks. The completed matrix lets us compare both average predicted retention and the rankings assigned to the same application records across firms.

These comparisons motivate the two questions organizing our analysis: whether each model predicts held-out retention in the firm on whose data it was trained, and whether these locally predictive models preserve a common ordering of applications across firms. We refer to these properties as local predictive validity and organizational portability, respectively.

\input{analysis/output/paper_evaluation/production/figure_application_firm_matrix.tex}

\section{Local Prediction and Organizational Portability}\label{sec-local-prediction-portability}

The application-by-firm matrix separates a question each firm can answer with its own data from portability questions that require multiple models from across a platform. An individual firm can ask whether its assessment predicts retention among its own employees. By applying every local model to the same production applications, the platform can additionally ask whether the models preserve a common application ordering and how far their predicted retention probabilities differ for the same profiles. The first question concerns whether a model works in the environment where it was estimated; the latter questions concern whether what it learns is portable across organizational environments.

This section develops the framework for these comparisons. We first define local predictive validity and explain why portability can be evaluated in two distinct ways. We then describe how we estimate comparable firm-specific retention models, establish their predictive performance within their own evaluation samples, and calibrate the probability scales used in the analysis. Finally, we apply the resulting models to the common production population and define the two portability comparisons used in the empirical analysis. Because the design does not observe alternative-firm retention outcomes for a common set of employees, these comparisons characterize the portability of fitted retention functions rather than the outcomes applicants would causally realize if assigned to another firm.

\subsection{Defining Local Validity and Portability}\label{defining-local-validity-and-portability}

We begin with the prediction problem faced by an individual firm. For each firm \(f\), define

\[
m_f(x) \equiv \Pr(Y_i=1 \mid X_i=x, F_i=f)
\]

where \(Y_i\) equals one if the worker associated with evaluation application \(i\) remains employed beyond 90 days, \(X_i\) is the application's common assessment profile, and \(F_i\) identifies the firm in which the retention outcome is observed. The function \(m_f\) describes how assessment profiles map into observed retention within firm \(f\).

Let \(\widehat{m}_f\) denote the corresponding fitted model. We assess its local predictive validity using held-out evaluation applications from the same firm. A model has local predictive validity when \(\widehat{m}_f(X_i)\) ranks applications associated with retained workers above those associated with workers who exit within 90 days. Local validity asks whether the assessment contains useful predictive information in the organizational environment where the model was estimated, and is therefore addressable by standard prediction metrics from the machine learning literature.

High local validity does not imply organizational portability. Even if \(\widehat{m}_f=m_f\), the local retention functions \(m_f\) and \(m_g\) need not agree. At the same time, differences in predicted retention levels do not necessarily imply disagreement about which profiles are relatively favorable: firms can differ in average retention and in the probabilities assigned to the same profiles while ranking applications identically. For example, the same application may receive a 60 percent predicted retention probability from a model estimated at a firm with high overall retention and a 40 percent probability from a model estimated at a lower-retention firm, even though it ranks at the median under both models.

We therefore define ordinal portability in terms of a common application ordering rather than equality of predicted probabilities. A common ordering is portable across firms if there exists a scalar score \(s(x)\) and strictly increasing firm-specific transformations \(h_f(\cdot)\) such that

\[
m_f(x)=h_f\!\left(s(x)\right)
\]

for every firm \(f\). The transformations allow firms to differ in average retention and in the probabilities assigned to the same profiles. Their monotonicity nevertheless requires firms to agree on application ordering: if one profile ranks above another at one firm, the same ordering must hold at every firm. The common score \(s(x)\) is thus a restriction on the structure of the local retention functions rather than a separately estimated prediction object. Portability can therefore fail even when every model has local predictive validity, through the lack of a common scoring function \(s(x)\).

We use the production application-by-firm prediction matrix to evaluate portability in two ways. We first test the ordinal restriction by comparing the percentile rankings assigned by the local models to the same applications. We then examine a stronger probability-level restriction by equalizing average predicted retention across models and measuring the profile-specific differences that remain. The latter comparison retains the probability scale rather than allowing arbitrary monotone firm-specific transformations. Both comparisons hold the assessment features, model class, and scored population fixed; only the firm-specific evaluation data used to estimate each local function vary.

\subsection{Estimating Comparable Local Models}\label{subsec-model-development}

Cross-firm disagreement is informative only if each model first predicts retention at its own firm. Otherwise, differences in scores could reflect estimation noise rather than organizational heterogeneity. We therefore establish local predictive validity before turning to portability.

For each firm \(f\), our access to the platform's underlying data allows us to estimate a separate retention function, \(\widehat{m}_f\), using only that firm's evaluation applications and observed 90-day outcomes. The models actually deployed by the platform and their prediction scores were not available to us. Our estimated models use the same 49 predictors: responses to the 42 common assessment items and the seven associated composite trait scores. Demographic characteristics enter neither our models nor the models used by these firms. The information set and estimation protocol are therefore common across firms; what varies is the firm-specific evaluation data from which each retention function is learned.

We separate estimator choice, performance assessment, and final model fitting. First, we compare five candidate estimator classes under a common three-fold, outcome-stratified design and select the class that performs best consistently across firms. Second, after fixing the estimator class, we use a separate nested cross-validation exercise to estimate local predictive validity. Hyperparameters are retuned within each outer-training sample, and the reserved outer fold is used only for assessment. Finally, we fit one model on each firm's full evaluation sample using the hyperparameters selected during candidate development. After fitting, we apply a monotone transformation, known as Platt calibration, that maps each model's outputs onto a probability scale estimated from out-of-fold predictions and realized outcomes in the corresponding evaluation sample. Appendix \ref{sec-predictive-model-assessment} describes the estimation, and subsequent calibration, in detail.

Candidate development compares logistic regression, random forests, polynomial support vector machines, gradient-boosted trees, and neural networks. We tune each flexible estimator within the common three-fold design; Appendix \ref{sec-predictive-model-assessment} reports the estimator-specific search procedures and complete comparison. The flexible estimators substantially outperform logistic regression, consistent with important nonlinearities and interactions in the relationship between assessment responses and retention. Random forests attain the highest candidate AUC at every firm, so we select them as the common estimator class for the analysis.

With the estimator class fixed, the nested assessment yields random-forest AUCs of 0.855 (95 percent CI: 0.837--0.873), 0.891 (0.880--0.902), and 0.914 (0.905--0.923) for Firms A, B, and C, respectively. These results establish our premise: the estimated models are meaningful summaries of each firm's learned retention function and have high local predictive validity.

\subsection{Cross-Scoring and Measures of Portability}\label{sec-cross-scoring}

Having established that each local model predicts retention within its own firm, we compare the fitted retention functions on a common population of production applications. Portability can be evaluated in two distinct ways. For a firm choosing among applications for a fixed number of positions, the relevant question is whether different models produce the same screening order. For a platform considering whether an application might be better matched to another firm, ordering alone is insufficient: the platform also needs to know by how many percentage points predicted retention would change under the alternative firm model. The first question is ordinal; the second requires retaining the probability scale.

For application \(i\), let \(\widehat{q}_{if}\) denote the probability of attrition within 90 days assigned by firm \(f\)'s model. The corresponding predicted retention score is

\[
\widehat{p}_{if}
\equiv 1-\widehat{q}_{if}
=\widehat{m}_f(X_i),
\]

where \(X_i\) is the application's common assessment profile. Let \(F_i\) denote the firm receiving application \(i\). When \(f=F_i\), \(\widehat{p}_{if}\) is an own-context prediction; when \(f\neq F_i\), it is a cross-scored or alternative-context prediction. Because every local model uses the same feature space \(X\), each model can score applications received by any of the three firms.

Applying all three models to every production application yields the application-by-firm prediction matrix introduced in Figure \ref{fig-application-firm-matrix}. This common-population matrix permits us to separate agreement about application ordering from agreement about predicted retention levels. These dimensions need not coincide. Two models can rank every application identically while assigning substantially different probabilities; conversely, they can reverse the ordering of closely scored applications while differing little in probability levels. Production applications do not have observed retention outcomes, so both measures compare fitted models on the same profiles rather than evaluating scores against realized outcomes.

\textbf{Ordinal portability.} Our first measure asks whether firm identity changes the relative ordering of applications. This is the relevant notion of portability for rank-based screening: if firms fill a fixed number of positions, differences in average retention or in the probabilities assigned to particular applications do not affect selection provided that every model produces the same ordering.

We convert each model's production-sample predictions into percentile rankings. Let \(r_{if}\) denote application \(i\)'s descending rank under firm \(f\)'s model among the \(N\) production applications, with tied predictions assigned their average rank. We define

\[
R_{if}
=
100\left(
1-\frac{r_{if}-1}{N-1}
\right),
\]

so that higher percentiles indicate stronger predicted retention. In the absence of estimation error, the common-ranking condition \(m_f(x)=h_f(s(x))\) implies

\[
R_{if}=R_{ig}
\qquad
\text{for every application }i
\text{ and every pair of firms }f,g.
\]

We assess this restriction using binned rank mappings between models. Exact ordinal portability implies a binned mapping that lies on the 45-degree line and highlights whether disagreement is concentrated in particular parts of the application distribution. This comparison reveals whether alternative firm models would reorder the screening queue, but deliberately discards the magnitude of the underlying probability differences.

\textbf{Predicted-retention differences and baseline equalization.} Our second measure retains the probability scale. For an application received by firm \(F_i\), the difference \(\widehat{p}_{if}-\widehat{p}_{iF_i}\) describes the model-implied change in predicted retention when that application is evaluated under firm \(f\)'s fitted retention function rather than the receiving firm's function. This is the score-based object used below to ask whether rerouting an application to another firm is predicted to improve retention and, if so, by how many percentage points.

Before baseline equalization, these probability differences combine two conceptually distinct sources. One model may assign higher predicted retention to nearly every application because its firm has a higher baseline retention environment. Alternatively, the models may evaluate the same assessment profile differently. To distinguish these possibilities, consider the stronger benchmark

\[
\operatorname{logit}m_f(x)=\alpha_f+g(x),
\]

where \(\alpha_f\) captures a firm-specific retention baseline and \(g(x)\) is a profile mapping common to all firms. Whereas ordinal portability permits arbitrary strictly increasing firm-specific transformations \(h_f(\cdot)\), this benchmark allows the fitted functions to differ only through an additive firm component on the log-odds scale. After removing that component, the models should assign the same predicted retention to every profile.

We operationalize this benchmark by equalizing average predicted retention across models. Let

\[
\eta_{if}
=
\log\left(\frac{\widehat{p}_{if}}{1-\widehat{p}_{if}}\right)
\]

and define the equalized prediction

\[
\widehat{p}_{if}^{\,\mathrm{eq}}
=
\frac{\exp(\eta_{if}+\delta_f)}
{1+\exp(\eta_{if}+\delta_f)},
\]

where \(\delta_f\) is chosen so that

\[
\frac{1}{N}\sum_{i=1}^{N}
\widehat{p}_{if}^{\,\mathrm{eq}}
=
p^*
\]

for every firm \(f\). In the baseline specification, \(p^*\) is the simple average of the three firms' evaluation-sample retention rates, giving each firm equal weight. Because this transformation is strictly monotone,

\[
R_{if}^{\,\mathrm{eq}}=R_{if},
\]

so equalization changes each model's average predicted level without changing its ordering of applications. Differences that remain after equalization cannot be accounted for by a firm-specific additive difference in log-odds baselines. Under this normalization, they summarize differences in the profile-specific structure of the fitted retention functions.

The two measures therefore serve different purposes. Rank comparisons reveal whether using another firm's model would change which applications are selected. Probability comparisons quantify the model-implied magnitude of those differences and allow us to separate average firm context from profile-specific disagreement. Neither exercise observes the same worker's retention under multiple firms. Alternative-context predictions, and the assignment improvements calculated from them, remain model-implied quantities rather than causal effects of rerouting an application.

\section{Evidence on Organizational Portability}\label{sec-portability-evidence}

\subsection{Ordinal Portability}\label{sec-common-ranking}

\begin{figure}[htb]

\centering{

\pandocbounded{\includegraphics[keepaspectratio]{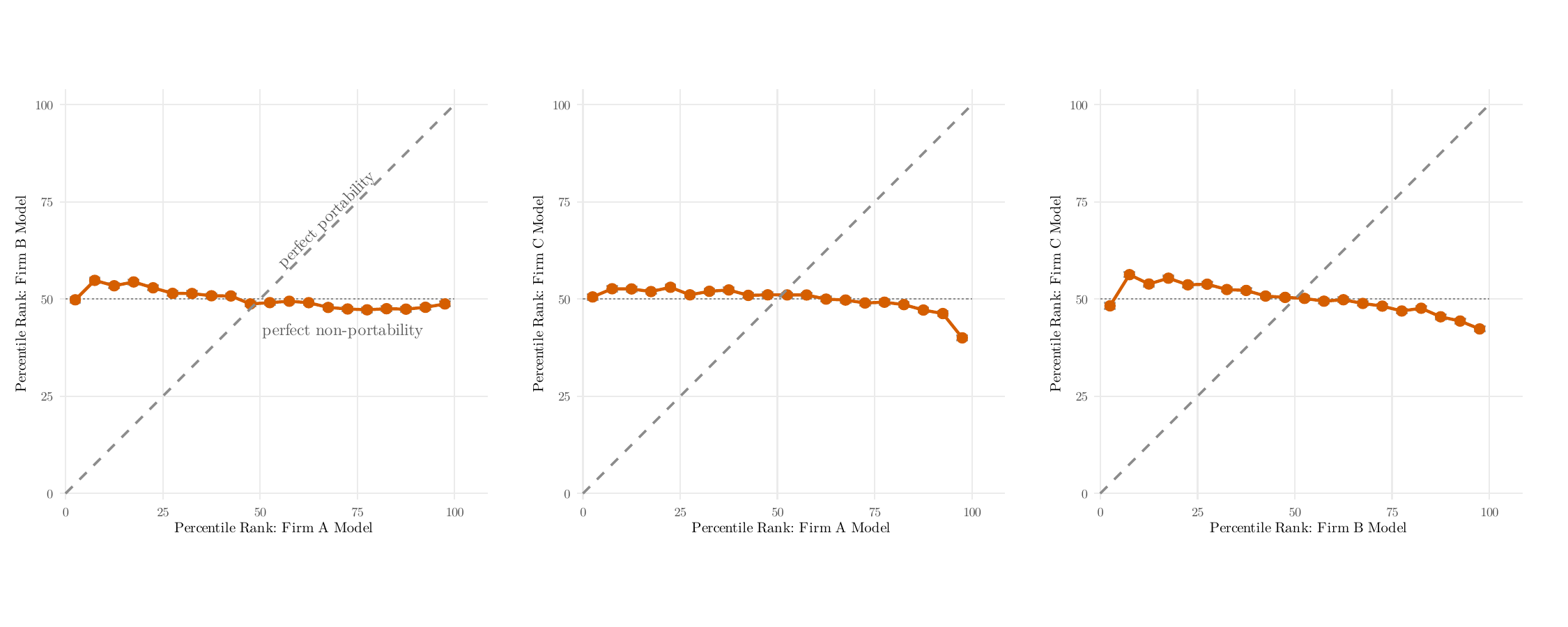}}

}

\caption{\label{fig-binned-percentiles}Binned cross-model percentile mappings in the production sample. Applications are grouped into five-percentage-point bins according to their percentile rank under the model on the horizontal axis. Points report the mean percentile rank under the model on the vertical axis; error bars report 95 percent confidence intervals for the bin mean. The dashed 45-degree line denotes the common-ordering benchmark.}

\end{figure}

We first use the common production population to examine ordinal portability. Figure \ref{fig-binned-percentiles} plots the binned rank mappings introduced in the previous section for each pair of firm models. Under a common ordering, applications placed in a given percentile range by one model would occupy the same range under another, and each mapping would follow the 45-degree line. This focus on ordinal rankings allows models to differ both in average predicted retention and in the probabilities assigned to individual applications.

All three mappings are approximately flat and lie far from the 45-degree benchmark. Across all three firm pairs, mean ranks under one model change little across percentile bins defined by the other. The Firm A--B mapping is closest to horizontal: as Firm A's ranking moves across the full distribution, the mean rank assigned by Firm B remains between the 47th and 55th percentiles. The A--C and B--C mappings likewise exhibit little systematic movement; to the extent that a trend is visible, it is weakly downward rather than upward. Moreover, the confidence intervals around the bin means are uniformly narrow---none spans more than 1.46 percentile points---so these shapes are estimated precisely within the production sample. The corresponding Spearman correlations for each pair are \(-0.067\) for Firms A and B, \(-0.078\) for A and C, and \(-0.106\) for B and C.

These results reveal that, despite the existence of a common assessment with highly predictive psychometric properties, retention-relevant rankings cannot be carried from one firm to another. A top application profile at Firm A is not necessarily the best for Firm B, nor vice versa. While these results are not indicative of whether a given application would cross screening thresholds at any given pair of firms (a decision dependent on an individual manager's hiring threshold and capacity at the time of application), they nonetheless reveal that cross-firm disagreement is not limited to different probability levels along a common ordering.

This result illustrates the model comparison externality. An individual firm observes only the ranking generated by its own retention data and therefore cannot determine whether that ranking is shared elsewhere. Because the platform administers a common assessment across firms, it can evaluate locally learned functions on the same applications and reveal disagreement that would otherwise remain hidden. For firms, this information bears on whether screening and recruitment rules can be transferred across settings, with direct implications for costly hiring and turnover. For policymakers, it helps distinguish a general shortage of suitable applicants from context-specific worker--firm fit, guiding whether scarce resources should be directed toward training or matching.

Appendix \ref{sec-ranking-stability} benchmarks this cross-firm pattern in two ways. Across repeated splits of each firm's evaluation sample, models trained on separate halves produce average production-sample rank correlations of \(0.376\), \(0.454\), and \(0.444\) for Firms A, B, and C, respectively. Rankings learned from different observations within the same firm are therefore substantially more aligned than rankings learned across firms. The appendix also reports a firm-label permutation test that reaches the same conclusion.

These results establish that firm-specific models would produce different screening queues from the same application pool. A ranking reversal does not, however, reveal how much is at stake. Two models may reorder applications whose predicted retention differs only trivially, or they may assign sharply different probabilities to the same profiles. This distinction matters for understanding the potential implications of rerouting applications. While rankings determine whom a firm would screen first, probability differences indicate whether evaluating an application under a different firm's retention function would materially change the predicted outcome. We therefore turn from whether the models disagree to the magnitude of that disagreement.

\subsection{Probability-Level Portability}\label{sec-probability-portability}

To measure the magnitude of disagreement, we compare each production application's predicted retention under the receiving-firm model with the highest prediction assigned by any of the three firm models. The resulting gap equals zero when the receiving-firm model assigns the highest score; otherwise, it measures how much higher the same application is scored under another firm's fitted retention function.

Firms' retention functions show considerable disagreement. Weighting receiving firms equally within assessment-index quintiles and quintiles by their production-sample shares, we find a mean pre-equalization gap of 21.8 percentage points. Under this weighting, the estimated shares of applications with gaps above 20 and 40 percentage points are 46.7 and 23.3 percent, respectively. This implies that the ordinal disagreement documented above is accompanied by substantial differences in the probability levels assigned to the same application records. Differences in ranking are driven by substantial differences in learned retention functions about each firm's retention outcomes.

These pre-equalization gaps do not, however, isolate disagreement about the mapping between assessment profiles and retention outcomes. A firm's model may assign systematically higher predictions because retention is generally higher in that firm's environment. The evaluation data, used in Section~\ref{sec-local-prediction-portability} to estimate these models, illustrate the importance of this distinction. In pooled logistic specifications using the seven composite traits, adding firm indicators increases pseudo-\(R^2\) by 0.052, whereas adding the traits conditional on firm indicators increases it by 0.002---a ratio of 21.6. This comparison concerns fit to observed retention rather than the decomposition of production-sample prediction gaps, and assumes a monotonic relationship unimposed on the random forest models. It nevertheless illustrates why model-wide level differences must be removed before the remaining variation can be interpreted as profile specific.

\begin{figure}[htb]

\centering{

\pandocbounded{\includegraphics[keepaspectratio]{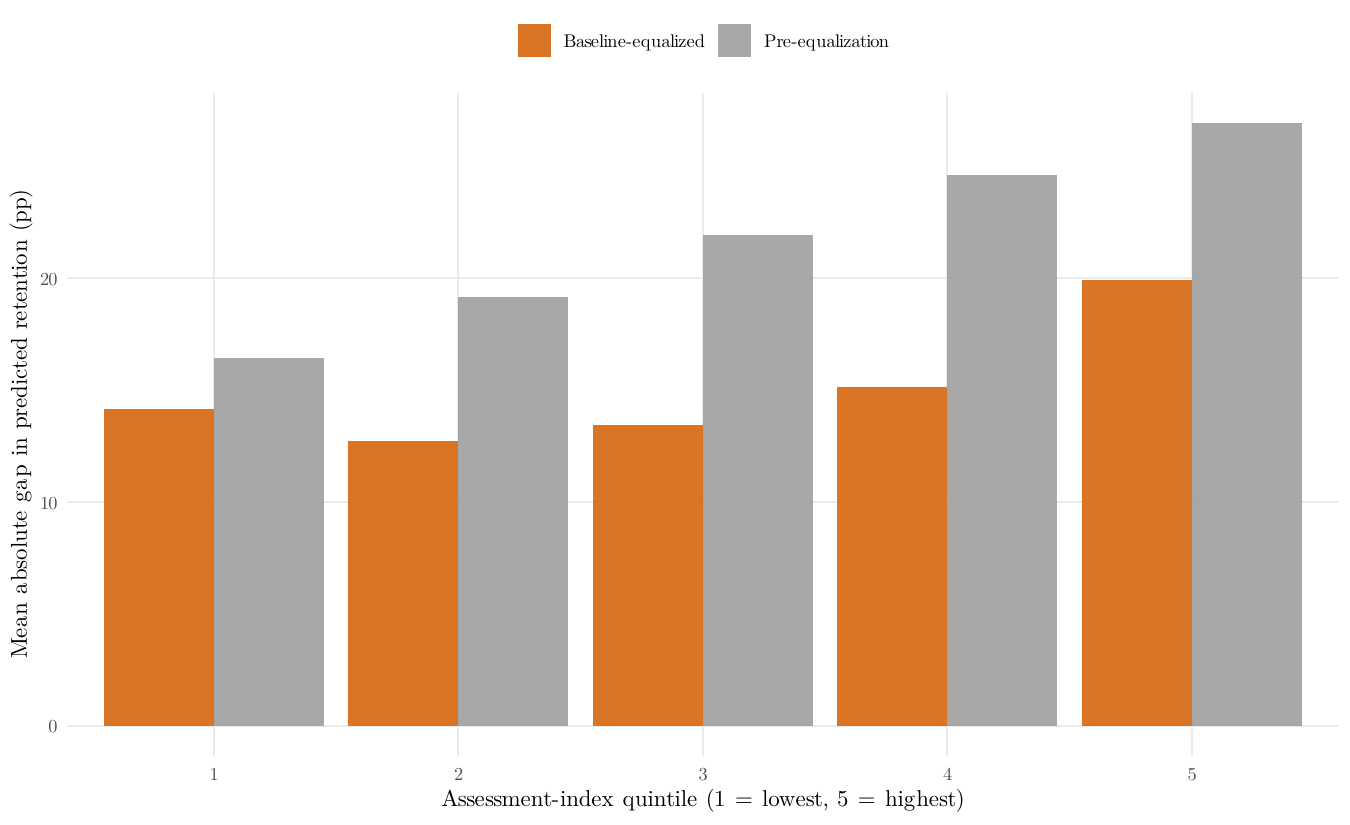}}

}

\caption{\label{fig-great-reversal}Pre-equalization and baseline-equalized prediction gaps by assessment-index quintile in the production sample. For each application, the gap is the difference between the highest predicted retention among the three firm models and the receiving-firm prediction. Estimates give each receiving firm equal weight within quintile. Baseline equalization uses additive log-odds shifts to bring each model to a common mean without changing its application ranking.}

\end{figure}

We therefore apply the baseline equalization introduced in Section~\ref{sec-local-prediction-portability}. The transformation shifts each model's predictions on the log-odds scale until all three have the same mean predicted retention. Because the transformation is strictly monotone, it preserves every model's application ranking and leaves the ordinal results in the previous subsection unchanged. Its purpose is narrower: to determine how much probability-level disagreement remains after removing differences attributable to model-wide retention levels. Figure \ref{fig-great-reversal} compares the pre-equalization and baseline-equalized gaps.

Baseline equalization reduces the mean gap from 21.8 to 15.1 percentage points. Differences in average predicted retention therefore account for about 30 percent of the pre-equalization gap. They do not account for the remainder: under the additive log-odds benchmark, the residual 15.1-point gap reflects differences in the profile-specific structure of the fitted retention functions. Average retention environments are thus important for probability comparisons, but they are not sufficient to reconcile the firm-specific models.

Figure \ref{fig-great-reversal} also shows where these differences occur within the assessment distribution. We divide production applications into quintiles of a pooled assessment index, constructed as the mean of the seven standardized composite trait scores. Pre-equalization gaps rise from 16.4 percentage points in the bottom quintile to 26.9 points in the top quintile. After baseline equalization, the gaps are relatively similar in the bottom three quintiles---14.2, 12.7, and 13.4 percentage points---before increasing to 15.1 points in the fourth quintile and 19.9 points in the fifth. Context-specific prediction differences are therefore especially pronounced among high-assessment profiles. This gradient provides suggestive evidence that the potential returns to interventions that improve matching may vary across applications and be greatest for high-index profiles. Thus, as prior experimental evidence suggests, interventions that reduce search frictions and improve worker--firm matching can affect labor-market outcomes \citep{Fernando2023, AbebeCariaFafchampsFalcoFranklinQuinnShilpi2025}, but their welfare and distributional consequences may be substantial and need not be evenly distributed across workers \citep{bassi2022}. Our results contribute to this literature by identifying a concrete empirical source of such heterogeneity: the model-implied scope for improved assignment rises sharply at the top of the assessment distribution.

Together, the ordinal and probability-level results establish two distinct forms of non-portability. The firm-specific models neither preserve a common application ordering nor reduce to a common profile mapping once their average predicted retention levels are equalized. The next section turns from the magnitude of this disagreement to its structure within the native fitted functions.

\section{The Structure of Cross-Firm Prediction Disagreement}\label{sec-model-explainability}

The preceding section shows that the three models assign different ranks and retention probabilities to the same production applications. That disagreement can arise at two different points. The models may rely more heavily on different assessment domains when distinguishing among applications. Even when the same domain matters, a given difference in its score may be associated with different retention differences across firms. The first possibility concerns what information drives the predictions; the second concerns how score differences relate to outcomes in each setting.

We examine these possibilities in sequence. We first ask which domains account for differences in predicted retention across applications. We then ask whether domain scores bear the same relationship to retention at each firm. The two questions are distinct: a domain can be important for prediction even when its association with retention does not differ across firms.

\subsection{Which domains account for firms' predictions?}\label{shared-information-in-the-fitted-scoring-rules}

The first question is which parts of the common assessment account for variation in each model's predictions. Each domain, however, enters the model through a composite score and six constituent responses. Examining these inputs separately would split the contribution of one measured trait across seven related variables. We therefore treat them as a single group and use grouped Shapley attributions, or grouped SHAP values, to measure how much each domain contributes to an application's prediction relative to the model's average \citep{LundbergLee2017, JullumRedelmeierAas2021}.

The decomposition begins with a model's mean prediction over the production applications. To build intuition, imagine revealing the seven domains to the model one at a time. A domain's contribution is the change in the prediction when that group is revealed. Because this change can depend on which domains have already been revealed, grouped SHAP averages the contribution across the possible orderings. Domains not yet revealed are drawn jointly from production records. The seven resulting contributions add to the difference between the application's prediction and the model's mean prediction.

Let \(\widehat p_{if}\) denote application \(i\)'s predicted retention under firm \(f\)'s fitted model, and let \(\mu_f\) denote that model's mean prediction over the \(N=172{,}242\) production applications. Exact grouped attributions satisfy

\[
\widehat p_{if}=\mu_f+\sum_{g=1}^{7}\phi_{ifg},
\qquad
\mu_f=\frac{1}{N}\sum_{i=1}^{N}\widehat p_{if}.
\]

All three models use the same production applications as the reference population, but each decomposition is centered on that model's own mean prediction. A positive \(\phi_{ifg}\) means that domain \(g\) contributes positively to application \(i\)'s deviation from that mean. It does not imply that increasing the domain score would increase retention. To compare predictive importance across domains, we use the mean absolute value of \(\phi_{ifg}\) across production applications.

\begin{figure}[H]

\centering{

\includegraphics[width=1\linewidth,height=\textheight,keepaspectratio]{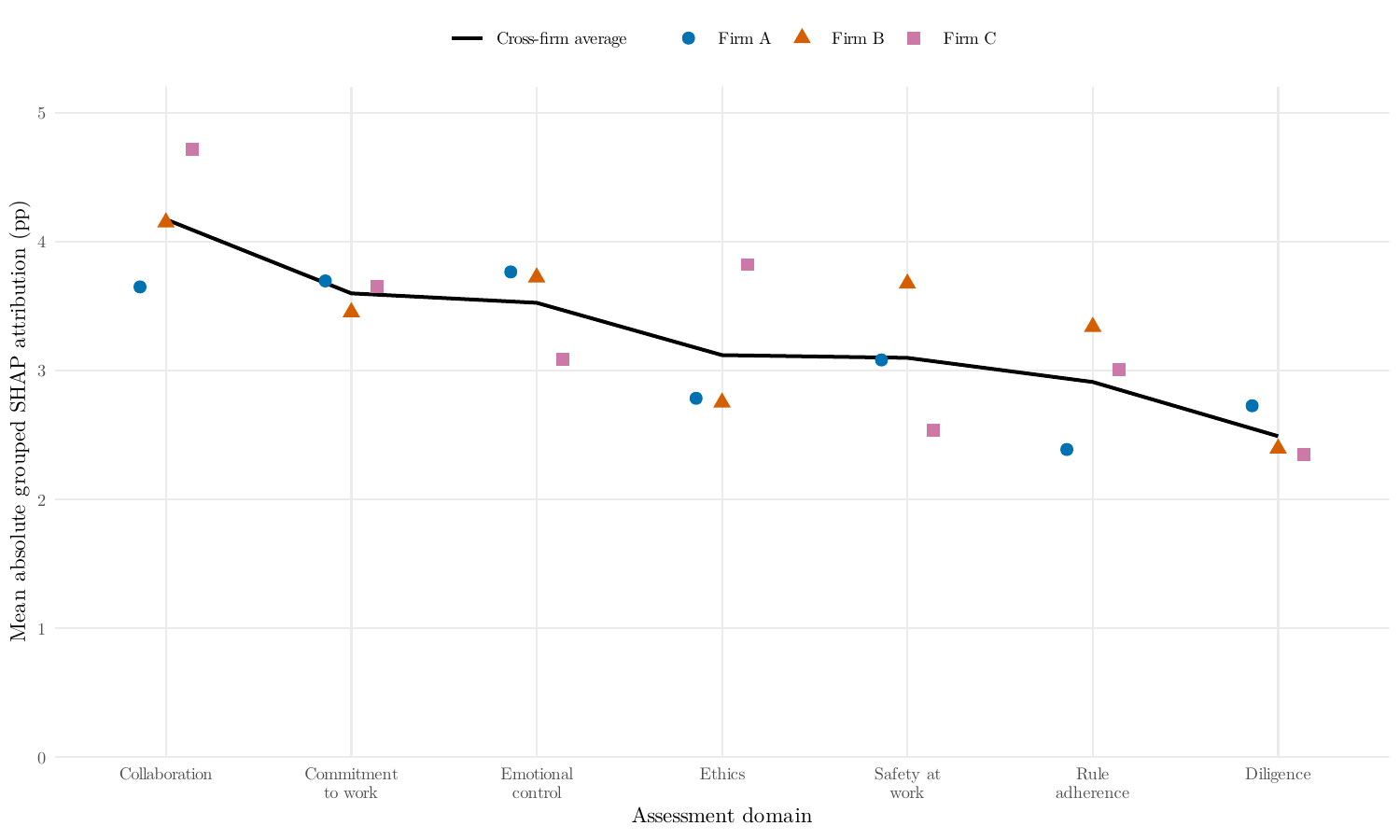}

}

\caption{\label{fig-group-shap-importance}Mean absolute grouped SHAP attributions by assessment domain and firm model in the production sample (\(N=172{,}242\)). Each point reports the mean absolute contribution of a domain to predicted 90-day retention relative to the corresponding model's mean prediction, measured in percentage points. The black line connects the unweighted mean across the three firm models within each domain. Domains are ordered by this cross-firm average. Each domain is treated as a single group containing its composite score and six constituent responses; attributions use 500 Monte Carlo draws per domain and application.}

\end{figure}

Figure~\ref{fig-group-shap-importance} shows broad overlap in the domains that contribute to the three models, but no common ordering of their importance. Collaboration has the largest mean absolute attribution when averaged across the three models: its mean absolute attribution is 3.65 percentage points for Firm A, 4.15 points for Firm B, and 4.72 points for Firm C. Emotional control, by contrast, has the largest mean absolute attribution under Firm A's model, at 3.76 points, but contributes only 3.09 points under Firm C's. The failure of portability therefore does not reflect three models built on wholly disjoint information. The common assessment captures domains that are relevant across firms without producing a common ordering of which domains matter most. For a firm, evidence that a domain is predictive elsewhere is therefore not enough to establish its relative importance locally.

Appendix Figure~\ref{fig-group-shap-importance-by-trait-quintile} repeats this unsigned comparison within quintiles of each domain's own score. These comparisons describe how the fitted models use the observed production applications. They do not decompose the baseline-equalized gaps in the preceding section, because each model retains its own probability scale and reference prediction. Nor do they ask how much predictive accuracy would be lost if a domain were unavailable. That question would require removing the domain and reestimating the model \citep{WilliamsonGilbertSimonCarone2023}.

\subsection{How do domain scores relate to retention across firms?}\label{a-common-comparison-in-the-underlying-retention-relationships}

A domain's contribution to existing predictions does not tell us whether a higher score has the same retention association across firms. The SHAP attribution explains why an application's existing prediction lies above or below the model's average; it does not compare conditional retention at different values of the domain score. We therefore turn to a common score comparison.

For each domain, we compare conditional retention at an application's observed mean item score with conditional retention at a score 0.25 higher, holding its other 36 item responses fixed. Because the recorded score has discrete support, this is a finite change rather than a derivative. We make the comparison for the same production applications at all three firms.

Making this comparison requires both samples, but in different ways. The evaluation applications contain both assessment responses and retention outcomes, allowing us to estimate the relationship between those responses and retention at each firm. The production applications contain no retention outcomes. They instead provide a common population of assessment profiles over which to compare the three firm-specific relationships. Each firm's contrast is therefore averaged over the same applications rather than over the composition of its own evaluation sample.

For domain \(g\), let \(R_g\) denote its six recorded responses, let \(S_g\) denote their mean, and let \(Z_g\) contain the other 36 responses, so that \(X=(R_g,Z_g)\). Let \(m_f(x)\) denote expected retention conditional on assessment responses in firm \(f\)'s evaluation population. For the production distribution \(Q\), define

\[
h_{fg}(s,z)=E_Q[m_f(R_g,z)\mid S_g=s,Z_g=z].
\]

A given domain mean can arise from several configurations of the six constituent responses. The function \(h_{fg}(s,z)\) averages Firm \(f\)'s conditional retention relationship over the configurations observed in the production population at domain mean \(s\) and other-domain profile \(z\). We compare \(h_{fg}(S_g+0.25,Z_g)\) with \(h_{fg}(S_g,Z_g)\). This construction holds the other 36 responses fixed without requiring every response in domain \(g\) to increase by 0.25 or imposing a particular path between the two domain means.

Because the domain mean ranges from zero to one, the comparison is defined for applications with \(S_g\le0.75\). We divide eligible applications into five groups according to their initial domain score, keeping identical scores together. The groups therefore approximate quintiles but need not contain equal numbers of applications. Within a domain, all three firms use the same applications and group boundaries; across domains, the eligible applications can differ. For initial-score group \(B_g=q\), the average retention difference is

\[
\Delta_{fgq}
=E_Q\!\left[
h_{fg}(S_g+0.25,Z_g)-h_{fg}(S_g,Z_g)
\mid S_g\le0.75,\ B_g=q
\right].
\]

Applications receive equal weight and remain in their initial group. The outcome regressions use the same 8,119 complete evaluation applications used to train the prediction models. The 172,242 production applications determine eligibility, initial-score groups, and the distribution of response configurations over which the contrast is averaged. We estimate \(\Delta_{fgq}\) and its sampling uncertainty using automatic debiased machine learning \citep{ChernozhukovNeweySingh2022}, adapted to combine the evaluation and production samples. Appendix \ref{sec-domain-target-details} describes the estimator, support conditions, and group construction.

The resulting estimates are conditional associations, not causal effects. A higher recorded score need not represent an improvement in the underlying trait, and the comparison does not identify the effect of training or reassignment. Interpreting the contrast for production applications also requires the conditional retention relationships estimated in the evaluation samples to carry over to the production population.

\begin{figure}[H]

\centering{

\includegraphics[width=1\linewidth,height=\textheight,keepaspectratio]{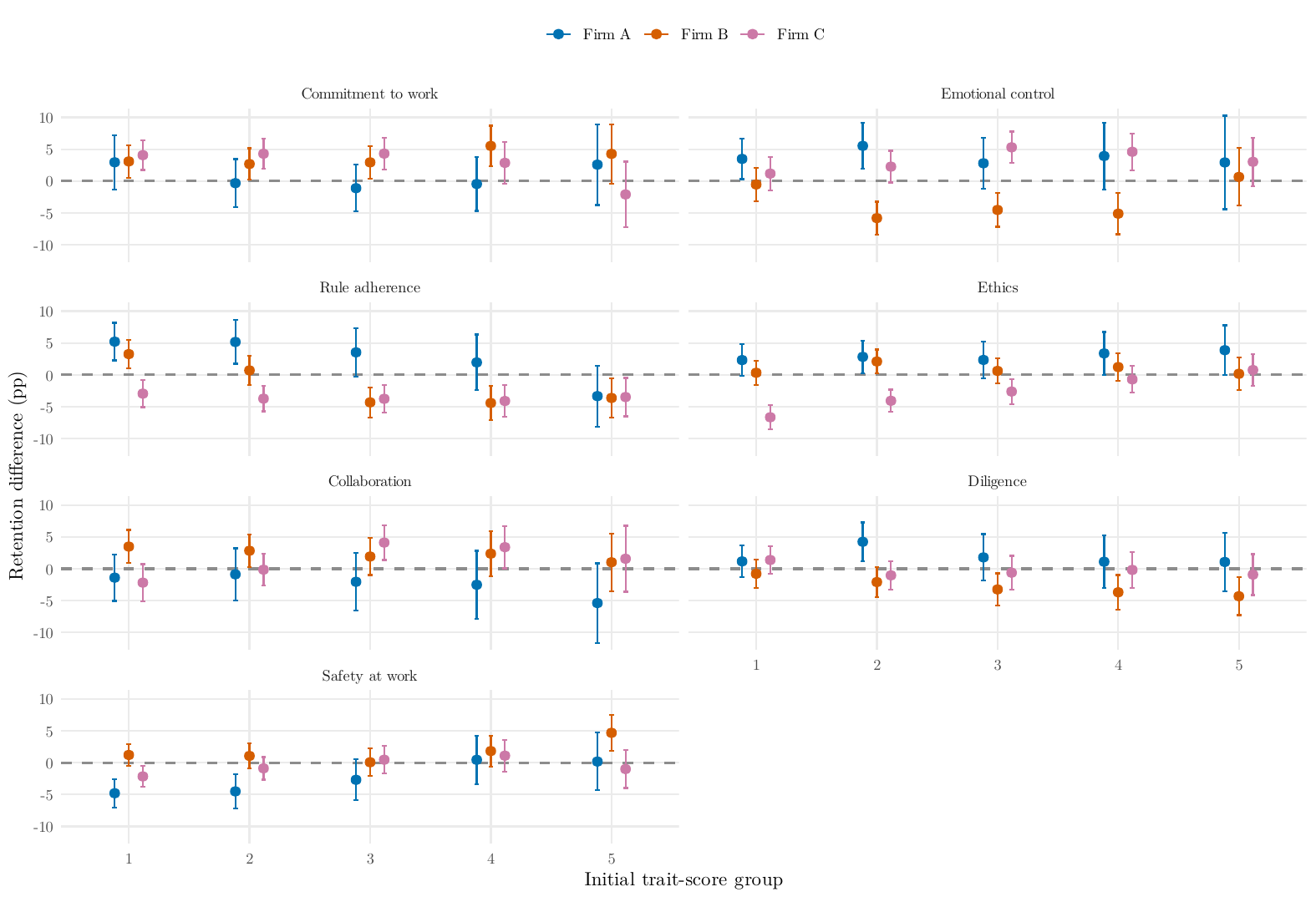}

}

\caption{\label{fig-domain-main}Firm-specific retention associations for a common domain-score increase, by assessment domain and initial domain-score group. Colors identify the three firms. Each point reports \(\Delta_{fgq}\) in percentage points: the average retention difference associated with a 0.25 increase in the domain's mean recorded score, holding the other 36 responses fixed. Within each domain, eligible production applications have initial scores no greater than 0.75 and are divided into five groups from lowest to highest initial score. Bars are pointwise 95 percent confidence intervals; the dashed line marks zero. Evaluation outcomes estimate the firm-specific retention relationships, while production applications define the common target population, groups, and response distribution. Support diagnostics flag 47 of the 105 estimates.}

\end{figure}

\protect\phantomsection\label{differences-across-firms} How do these score--retention relationships vary across firms? Figure~\ref{fig-domain-main} plots \(\Delta_{fgq}\) separately for each firm, using a different color for each firm. The points show the direction and magnitude of the retention association across the five initial-score groups; the bars are pointwise confidence intervals. Overlap between these intervals does not directly test whether two firms differ. We therefore base the cross-firm comparisons below on the pairwise differences \(\Delta_{fgq}-\Delta_{hgq}\) and their simultaneous confidence intervals, reported in Appendix Figure~\ref{fig-domain-pairwise}.

Emotional control provides the clearest evidence of cross-firm heterogeneity. Across the second through fourth initial-score groups, Firm B's retention contrast is 8.07 to 9.82 percentage points smaller than Firm C's; all three simultaneous 95 percent confidence intervals exclude zero, and none triggers the support warning described below. Across all domains, thirteen of the 105 simultaneous intervals exclude zero, with the differences concentrated in emotional control, rule adherence, ethics, and safety. Every interval for collaboration, commitment, and diligence includes zero. These estimates provide no comparable evidence of cross-firm differences for those three domains under the specified contrast, but they do not establish equality.

The remaining intervals that exclude zero warrant greater caution. The support diagnostic flags 47 of the 105 firm-specific estimates. These estimates enter 76 of the 105 pairwise comparisons, including nine of the thirteen whose simultaneous intervals exclude zero. For example, the Firm A--Firm B difference for safety in the first initial-score group is −6.00 percentage points, with an interval of {[}−10.87, −1.13{]}, but the comparison is flagged. A flag does not indicate an empty group or an undefined estimate. It indicates that a combination of response variables has effectively no variation in at least one fitting sample, limiting empirical support for the comparison. The appendix reports the diagnostic and examines the stability of the estimated group cutoffs. Passing this diagnostic addresses that specific support problem; it does not establish the remaining inference conditions.

Taken together, the estimates show that the relationship between a common score change and retention can differ across firms. Statistically significant differences appear in four of the seven domains---emotional control, rule adherence, ethics, and safety. This result does not establish equality elsewhere. It instead narrows where the data provide affirmative evidence that the score--retention relationship differs across firms.

Collaboration illustrates why the two exercises need not yield the same conclusion. Averaged across the three fitted models, it has the largest mean absolute attribution, yet none of its pairwise score-contrast intervals excludes zero. High SHAP importance means that observed collaboration responses account for substantial variation in existing predictions. It does not imply that a 0.25 increase has a nonzero average retention association or that this association differs across firms. The two approaches are therefore complementary rather than robustness checks on one another. Neither identifies firms' preferences, the causal sources of cross-firm differences, or the returns to changing workers' traits.

Together, the two exercises refine the earlier evidence of non-portability. Substantial cross-firm prediction disagreement coexists with a shared set of informative domains and selective evidence that particular score--retention relationships differ. A common assessment can therefore contain information that is useful at every firm without supporting a common screening rule. The results do not attribute the overall prediction gaps to any one domain; they show that common measurement and common use are distinct empirical properties.

This distinction is observed because this platform captures the model comparison externality. An individual firm can use its own outcomes to learn which assessment domains predict retention locally, but it cannot determine whether their relative importance or their relationship with retention is shared elsewhere. By linking common measurements to firm-specific outcomes, a multi-firm platform can reveal which parts of the signal are broadly shared and where the evidence is context specific. For firms, this information bears on whether a screening rule learned elsewhere is likely to travel or instead requires local validation and reestimation. For policymakers, it helps distinguish a common shortage of retention-relevant traits from a matching problem in which the same traits relate differently to outcomes across firms. The estimates do not identify the returns to training or reassignment, but they identify distinctions that single-firm data cannot recover.

\section{Conclusion}\label{sec-conclusion}

Platform provide a unique ability to capture model comparison externalities. They do so by hosting multiple models, each from different decision makers, all using standardized attributes and outcomes. Through this, platforms' insights extend beyond the specific data they observe: they can compare models used for decision making and give real answers about how market participants would respond to new circumstances under a market's current equilibrium. This externality can only exist because platforms are both hosts to these multiple models and coordination devices, ensuring these models use a common set of variables permitting such comparisons. If the same models exist but are siloed inside the servers of individual decision makers, the externality from comparing these models would not be captured. These comparisons generate insights about how decision making differs across economic actors, and how those differences might be impacted by policies that reallocate resources. These insights are valuable not only to the platforms, but also to firms and policymakers. 

We study this externality in the context of personnel selection. To do so, we partner with a AI hiring platform in Mexico, used by large employers of low-skill workers. For the three clients in our sample, this platform uses a standardized psychometric assessment to estimate firm-specific models of the 90-day retention likelihood associated with each job application. We use the models from these three firms, and the 172,242 job applications they received since using the platform for hiring, to create a firm-applicant matrix. This matrix creates a retention likelihood score for each application at each firm, not just the firm an application was submitted to. The ability to study how the same firms respond to job applications submitted to different firms distills the insights made possible by this model comparison externality.  

The models estimated from this data perform strongly on held-out data from their own firms, with AUCs ranging from 0.855 to 0.914. However, the insights about how to judge applications are not transferrable across firms: each firm's model is no better than chance when predicting retention likelihood at a different firm, based on the AUCs.  The relevant question is therefore not whether the assessment contains predictive information, but whether relationships that are useful within one firm remain useful in the same way elsewhere. Evaluating models by how they rank applications yields the same insight: an application's percentile rank according to one model is entirely uncorrelated with their percentile rank by a different firms model.  

This disagreement is not just over ranking. Hypothetically, models with only small disagreements about each applications' retention likelihood could produce substantive ranking disagreements. To test this, we consider which firm would score each application with the highest retention likelihood. We subtract the retention likelihood from where the application were actually submitted from this highest likelihood. We find that firms' models show substantial disagreement about the percentage likelihood of retention and this disagreement is heavily influenced by psychometric traits. For applications with average trait scores in the top quintile, the highest prediction from the three models exceeds the receiving-firm prediction by an average of 26.9 percentage points. For those in the bottom quintile, the gain was only 16.4 percentage points. 

We show that this disagreement is due to two distinct mechanisms. While part of this disagreement is due to worker-firm complementarities, the remainder is related to a recently documented mechanism known as sorting without complementarities. The latter arises because some firms simply have higher retention, regardless of the psychometric profile. We account for this mechanism by adapting the share-matching inversion method used in the discrete choice estimation literature. This allows us to equalize average retention across models, while preserving log-odds ratios within each model. This recalibration dropped gains in the top quintile by 7 percentage points and the bottom quintile by only 2.2 percentage points. These drops indicate that sorting without complementarities is present, as predicted by theory, are heterogeneously relevant. These findings are important for either platforms or policymakers trying to develop matching policies. They indicate policies emphasizing different mechanisms are important for different segments of the labor market. 

Finally, using methods from the explainable machine learning literature, we investigate the reasons behind these application-level disagreements. We find that firms' models emphasize, relatively, a common set of traits. However, they do not use those traits in the same way. For example, emotional control is generally relevant across firms. Yet, firms B and C have sharply different associations between emotional control and retention likelihood: firm B generally associates higher emotional control scores as decreasing retention likelihood at the margin, while firm C associates them with higher retention likelihood. Statistically significant differences also appear for ethics and rule adherence. The same assessment can therefore be informative across firms without yielding a screening rule that can simply be transferred among them.

Insights from these comparisons are valuable to firms, platforms, and policymakers. An individual firm can use its own model to establish whether an assessment predicts retention locally. However, access to information about differences across models can inform recruitment strategies. For example, our results overturn prevailing wisdom that firms are seeking a common, ideal profile in the low-skill labor market where the primary goal is high retention. Instead, different firms should be targeting recruitment resources towards different segments of the population. For platforms, these results show the potential for improved matchmaking between job applicants and firms. Further, they highlight the importance of carefully tailoring matching mechanisms: frictions have heterogeneous relevance across the psychometric trait distribution. Policymakers interested in improving their local labor markets learn that matchmaking and skill development programs are investments with distinct impacts across the distribution of existing trait distribution. Finally, these insights inform policymakers about the  value of allowing large platforms. These model comparison externalities are much weaker in a fragmented landscape with many platforms. Consequently, as in the case of traditional network externalities, antitrust concerns must be balanced with the concern of not capturing the model comparison externalities illustrated in this research. 

Our analysis does not causally establish whether directing an applicant to another firm would improve retention; answering that question requires experimental variation in assignment. It does establish a distinct source of value from platforms. By linking models used by distinct decision makers for making real world decisions, it can produce insights about how different market participants would react differently to the same circumstances. This information cannot be recovered from isolated client data; it requires platforms that make cross-firm comparison possible.

Looking ahead, because knowledge and preferences are dispersed and context-specific \citep{Hayek1945}, AI is likely to look less like a single model deciding how to allocate resources and more like an economy of many local models, from agents that shop for individual consumers to firm-specific models that manage each firm's hiring. If policymakers can access these models, they can simulate policy alternatives using the actual response functions of market participants rather than a central model's guess at them. Our results offer an early example: access to multiple firms' hiring models reveals how workers could be better allocated across firms. AI may thus make centralized calculations over a decentralized economy feasible, not by gathering dispersed knowledge in one place but by aggregating the local models that embed it---a kind of Hayek in silico.

\section{References}\label{references}

\renewcommand{\bibsection}{}
\bibliography{references.bib}

\appendix

\section{Predictive Model Assessment}\label{sec-predictive-model-assessment}

This appendix documents the construction, local assessment, and probability calibration of the three firm-specific random forest models used in the paper. It separates four stages: candidate development, held-out validation, final full-sample fitting, and probability calibration. Candidate development compares estimator classes and supplies the random-forest settings carried into the final models. A separate nested cross-validation exercise retunes temporary models within each outer-training sample solely to measure held-out predictive fit. We then fit the three production models using the carried-forward settings and calibrate their outputs using held-out predictions from the corresponding evaluation samples.

\subsection{Prediction Task and Common Estimator}\label{prediction-task-and-common-estimator}

For each evaluation application \(i\), the prediction target is the indicator \(D_i = \mathbf{1}[\text{Tenure}_i \leq 90]\). A firm-specific random forest produces an initial estimate of early-attrition risk, \(\widehat{q}^{RF}_{if}=\Pr(D_i=1\mid X_i,F_i=f)\), with corresponding uncalibrated retention prediction \(\widehat{p}^{RF}_{if}=1-\widehat{q}^{RF}_{if}\). The calibration procedure below maps \(\widehat{p}^{RF}_{if}\) into the probability \(\widehat{p}_{if}\) used in the main analysis. The evaluation sample contains 8,119 applications: 1,682 from Firm A, 3,123 from Firm B, and 3,314 from Firm C. Each model uses the same 49 predictors: responses to 42 common assessment items and seven composite trait scores. Demographics do not enter the models. Each firm-specific model uses only applications and realized outcomes from its corresponding firm.

Before fixing the estimator class, we compare logistic regression, random forests, polynomial support vector machines, gradient-boosted trees, and neural networks under a common three-fold, outcome-stratified design. Logistic regression has no tuning parameters. For each flexible estimator, three independent searches begin with a space-filling design and then use Bayesian optimization. The polynomial-SVM search allows degrees one through ten and tunes the cost and scale parameters. In every case, hyperparameters are chosen by mean AUC across the three folds; for the polynomial SVM, we select the highest-AUC configuration that produces predictions and finite AUC in all three folds when refit.

Random forests attain the highest candidate AUC at every firm: 0.858, 0.892, and 0.916 for Firms A, B, and C. Their advantage over the next-best candidate is 0.032, 0.045, and 0.035, respectively. The candidate comparison therefore selects random forests as the common estimator class, allowing us to hold both the estimator and the downstream explanation procedure fixed across firms.

\input{analysis/output/paper_evaluation/candidate_model_comparison/table_candidate_model_auc.tex}

Within each resample, the preprocessing recipe removes zero-variance predictors, median-imputes missing numeric inputs, and normalizes numeric inputs using parameters estimated only from that resample's training observations. The nested cross-validation exercise below supplies the evidence on local predictive validity; neither the candidate comparison nor the common estimator choice does so by itself.

\subsection{Nested Validation and Final-Model Settings}\label{nested-validation-and-final-model-settings}

We separate model tuning from performance assessment using nested cross-validation. For each firm-specific model, the outer loop partitions evaluation applications into three folds stratified by the early-attrition outcome. For each outer fold, we reserve one fold for assessment and conduct the entire tuning procedure using only the remaining two folds. Within that outer-training sample, a second set of three stratified folds supplies the validation outcomes used to choose hyperparameters. The reserved outer fold is scored only after preprocessing and tuning are complete.

The optimization criterion is inner-fold AUC. We tune three random-forest parameters: the number of candidate inputs considered at each split (\texttt{mtry}, from 1 to 49), the number of trees (\texttt{trees}, from 500 to 3,000), and the minimum terminal-node size (\texttt{min\_n}, from 1 to 30). Each search begins with a 60-point space-filling design. A Gaussian-process surrogate then maps previously evaluated parameter combinations into expected AUC, and an expected-improvement acquisition rule chooses up to 150 additional combinations. The search stops early after ten iterations without improvement.

Because this procedure is repeated within each outer-training sample, the selected parameters vary across folds. Across the three outer folds, the selected ranges are 3--22 for \texttt{mtry}, 1,432--2,974 trees, and 1 for \texttt{min\_n} for Firm A; 7--13, 1,891--2,756, and 1 for Firm B; and 4--12, 947--2,039, and 1 for Firm C.

For production cross-scoring, we take the random-forest parameter rows selected by the candidate-development search---there is no second tuning pass on the complete samples---and fit one model to each firm's full evaluation sample. The final settings for \texttt{mtry}, \texttt{trees}, and \texttt{min\_n}, respectively, are 3, 547, and 2 for Firm A; 16, 2,469, and 1 for Firm B; and 29, 808, and 2 for Firm C. These full-sample models are distinct from the fold-specific validation models and are the only fitted models used for production cross-scoring and downstream substantive results.

\input{analysis/output/paper_evaluation/manuscript/tables/table_final_rf_hyperparameters.tex}

\subsection{Local Out-of-Sample Fit}\label{local-out-of-sample-fit}

Table \ref{tab:paper-evaluation-nested-cv} asks whether each model discriminates between early attrition and retention on applications excluded from both preprocessing and hyperparameter selection. It reports AUC from the concatenated predictions for the three held-out outer folds. AUC is the relevant fit measure because the paper's screening comparisons depend on application rankings rather than on a single probability cutoff. Firm A's AUC of 0.855 means that the fixed random-forest class assigns higher early-attrition risk to a randomly selected evaluation application with attrition within 90 days than to one with retention beyond 90 days in 85.5 percent of such pairs.

\input{analysis/output/paper_evaluation/manuscript/tables/table_nested_cv_firm_models.tex}

For the firm-specific random forests, out-of-fold AUC is 0.855 for Firm A, 0.891 for Firm B, and 0.914 for Firm C. The corresponding 95 percent confidence intervals are {[}0.837, 0.873{]}, {[}0.880, 0.902{]}, and {[}0.905, 0.923{]}. Even the lower confidence bound for Firm A remains well above the 0.50 chance benchmark. The exercise therefore establishes local predictive validity under application-level resampling: no application contributes to preprocessing, tuning, or estimation for the outer fold in which its prediction is evaluated. It does not, by itself, establish calibration or cross-firm portability, and its temporary fitted models are not reused elsewhere in the analysis.

\subsection{Probability Calibration}\label{probability-calibration}

Because AUC depends only on ranking, strong discrimination does not ensure that a model's reported probabilities correspond to observed retention rates. We therefore estimate a separate Platt transformation for each firm. Let \(\widehat p^{RF}_{if}\) denote the uncalibrated random-forest prediction of retention. The calibrated prediction is

\[
\widehat p_{if}
=
\operatorname{logit}^{-1}
\left[
\widehat\alpha_f
+
\widehat\beta_f
\operatorname{logit}\left(\widehat p^{RF}_{if}\right)
\right],
\]

where \((\widehat\alpha_f,\widehat\beta_f)\) are estimated from realized retention and out-of-fold predictions in firm \(f\)'s evaluation sample. Each out-of-fold prediction is generated by a model estimated on the other four folds using the final hyperparameter settings.

All calibration parameters are estimated in the evaluation sample, where 90-day retention is observed. To assess the calibration procedure out of sample, we divide the out-of-fold random-forest predictions into five calibration folds, estimate \((\alpha_f,\beta_f)\) on four folds, and evaluate the transformation on the fifth. We then estimate one final transformation for each firm using all of its out-of-fold evaluation predictions and observed outcomes. That fixed transformation is applied, without further estimation, to the predictions generated for production applications by the random forest fitted on the full evaluation sample.

\begin{figure}[htb]

\centering{

\pandocbounded{\includegraphics[keepaspectratio]{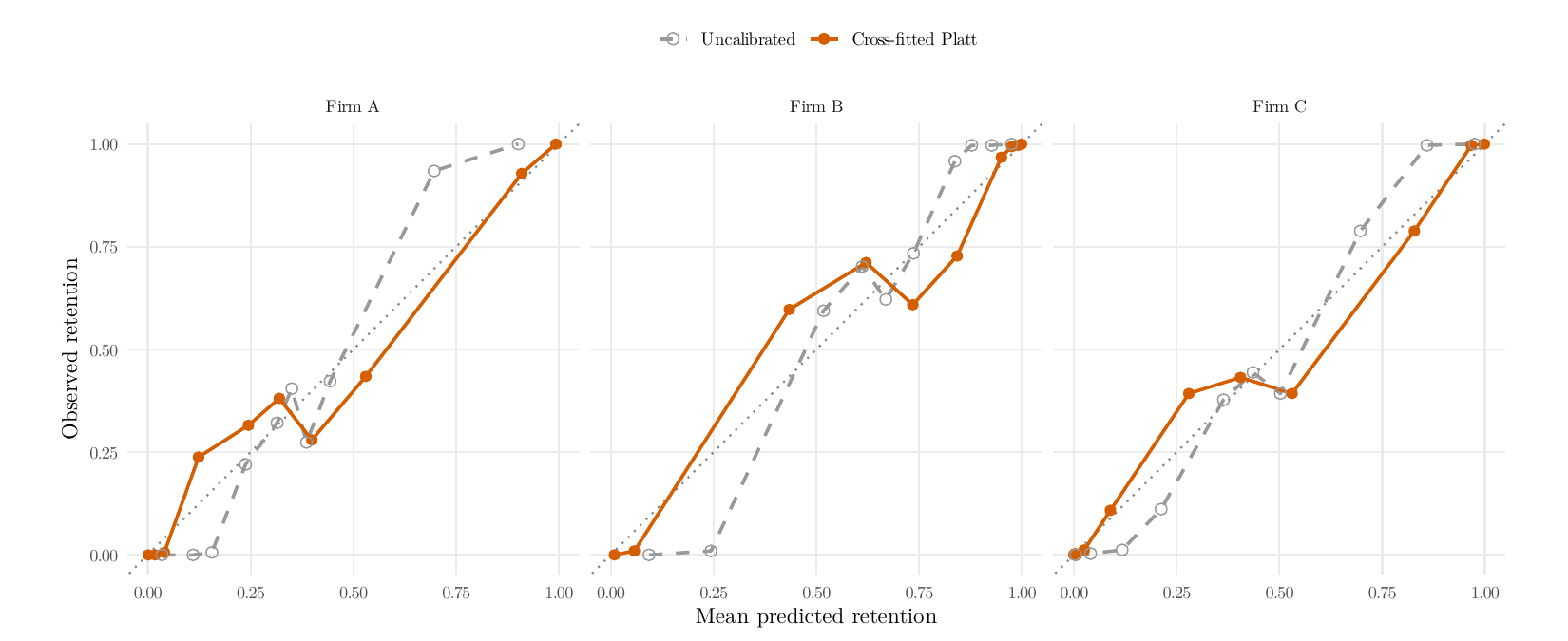}}

}

\caption{\label{fig-calibration-reliability}Reliability diagrams before and after Platt scaling for each firm-specific model. The diagonal denotes perfect calibration.}

\end{figure}

Figure~\ref{fig-calibration-reliability} compares observed retention with mean predicted retention across deciles of the out-of-fold prediction distribution, before and after cross-fitted Platt scaling. Before scaling, the average absolute gap between predicted and observed retention across these deciles is 8.4 percentage points for Firm A, 8.8 points for Firm B, and 6.3 points for Firm C. After scaling, the corresponding gaps fall to 5.4, 5.9, and 3.8 percentage points. Platt scaling therefore brings predicted retention closer to observed retention in each firm's evaluation sample.

\section{Directed Cross-Firm Transfer Performance}\label{sec-directed-transfer}

As a supplementary evaluation-sample exercise, we ask whether a model estimated at one firm retains predictive ordering for outcomes observed at another. For each ordered pair of firms \(f\) and \(g\), we calculate

\[
\operatorname{AUC}_{f\rightarrow g}
=
\operatorname{AUC}
\left(
\left\{\widehat{p}_{if},Y_i\right\}_{i:F_i=g}
\right),
\]

where \(Y_i\) indicates observed 90-day retention. For \(f=g\), the diagonal cells use the held-out predictions from the nested local-validity exercise. For \(f\neq g\), the off-diagonal cells apply the final model estimated at firm \(f\) to the evaluation applications observed at firm \(g\). The empirical benchmark in an off-diagonal cell is therefore the outcome observed at \(g\), not the ranking produced by \(g\)'s model.

Applying each final model to the other firms' evaluation samples produces six transfer AUCs between 0.473 and 0.514 (Table \ref{tbl-cross-firm-auc}). Five of the six confidence intervals include 0.50; the remaining estimate, B-to-C, is below chance (0.473, 95\% CI {[}0.453, 0.493{]}). Taken together, these estimates provide little evidence that a model learned at one firm supplies a stable outcome-referenced ordering at another. Because AUC depends only on ordering, the exercise does not test equality of probability levels or calibration.

\input{analysis/output/paper_evaluation/manuscript/tables/table_r2_cross_firm_auc.tex}

\section{Benchmarks for Cross-Firm Ranking Disagreement}\label{sec-ranking-stability}

Two checks address whether the observed mean pairwise cross-firm rank correlation of \(\bar{\rho}=-0.084\) reflects systematic model disagreement or statistical artifacts.

\subsection{Same-Context Stability}\label{same-context-stability}

We benchmark cross-firm ranking disagreement against the instability generated by estimating the same prediction problem on different observations. We randomly split each firm's evaluation applications into two halves, train separate models, and cross-score the production sample.

Across 100 splits per firm, within-firm rank correlations are 0.376 for Firm A (p10--p90: 0.311--0.433), 0.454 for Firm B (0.407--0.499), and 0.444 for Firm C (0.398--0.493). Mean out-of-fold AUCs for the two halves are 0.787/0.789, 0.804/0.803, and 0.847/0.845, respectively. These within-firm correlations are far above the observed cross-firm correlations (-0.067, -0.078, and -0.106). The exercise provides a same-context stability benchmark; it does not identify the organizational source of cross-firm disagreement.

\begin{figure}[htb]

\centering{

\pandocbounded{\includegraphics[keepaspectratio]{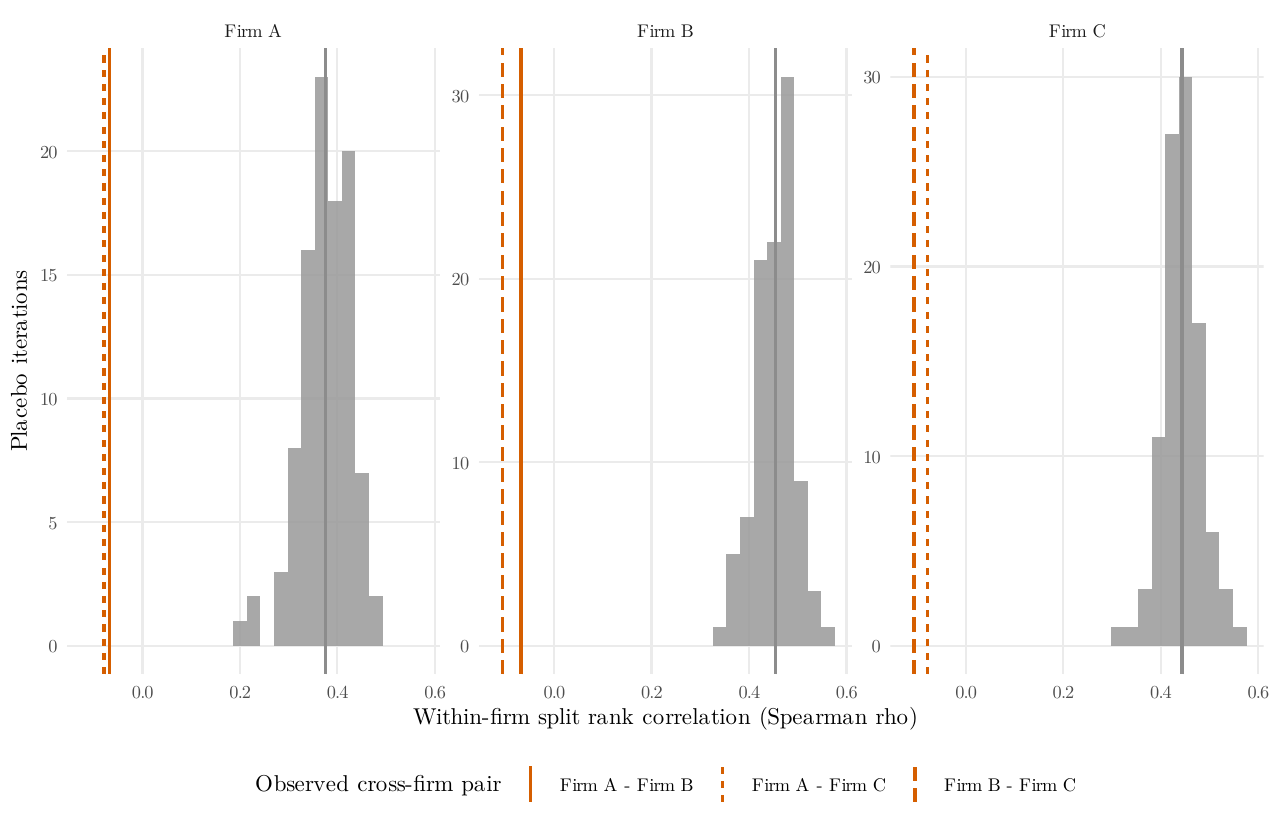}}

}

\caption{\label{fig-placebo-rho}Placebo split-firm rank correlation distribution. Bars report the number of split-specific \(\hat{\rho}\) estimates in each histogram bin across 100 random 50/50 splits of that firm's evaluation data, with separate models trained on each half. Dashed vertical lines mark observed cross-firm correlations.}

\end{figure}%

\subsection{Firm-Label Permutation Test}\label{firm-label-permutation-test}

We randomly permute firm labels 500 times, preserve firm sample sizes, and retrain the three firm-specific models for each permutation. The null distribution has mean 0.255 and SD 0.0199. The observed mean correlation of -0.084 is 17.00 standard deviations below that mean. No permutation is as small as the observed statistic; the prespecified one-sided Monte Carlo test with the plus-one correction gives \(p=0.002\).

\begin{figure}[htb]

\centering{

\pandocbounded{\includegraphics[keepaspectratio]{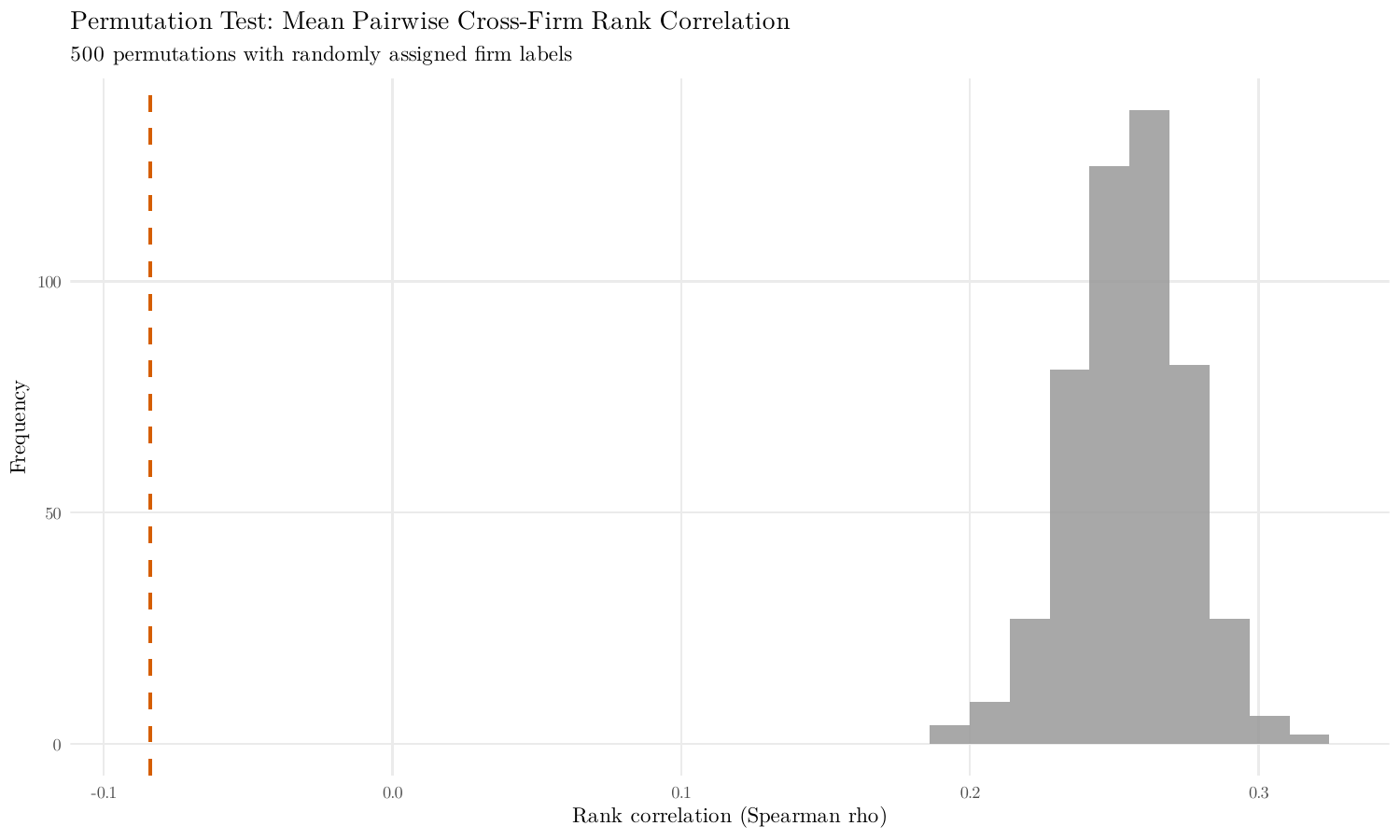}}

}

\caption{\label{fig-permutation-null}Permutation null distribution for the cross-firm rank correlation. Bars report the number of permuted mean pairwise rank correlations in each histogram bin across 500 random reshufflings of firm labels. The dashed vertical line marks the observed mean pairwise \(\bar{\rho}=-0.084\).}

\end{figure}%

\section{Psychometric Assessment}\label{sec-psych-assessment}

The psychometric instrument used in this study is a proprietary behavioral self-report inventory developed by the platform, administered to applicants via a WhatsApp chatbot at the time of application. The full instrument used in our analysis consists of 42 items organized into seven trait sections of six items each. Each item presents a statement about workplace behavior or attitude; respondents indicate how well the statement describes them using a scaled response. Item responses are normalized to \([0,1]\) and summed within each trait to produce a composite score ranging from 0 to 6. These seven composite scores, along with the underlying item-level responses, serve as the features in our predictive models.

The seven traits and illustrative examples of the types of statements in each section are summarized in Table~\ref{tbl-eri-traits} below. To protect the proprietary content of the instrument, example items are representative of each construct rather than verbatim survey questions.

\begin{longtable}[]{@{}
  >{\raggedright\arraybackslash}p{(\linewidth - 4\tabcolsep) * \real{0.3333}}
  >{\raggedright\arraybackslash}p{(\linewidth - 4\tabcolsep) * \real{0.3333}}
  >{\raggedright\arraybackslash}p{(\linewidth - 4\tabcolsep) * \real{0.3333}}@{}}
\caption{Employee Reliability Inventory: Trait Constructs and Illustrative Items}\label{tbl-eri-traits}\tabularnewline
\toprule\noalign{}
\begin{minipage}[b]{\linewidth}\raggedright
Trait
\end{minipage} & \begin{minipage}[b]{\linewidth}\raggedright
Construct
\end{minipage} & \begin{minipage}[b]{\linewidth}\raggedright
Illustrative Item
\end{minipage} \\
\midrule\noalign{}
\endfirsthead
\toprule\noalign{}
\begin{minipage}[b]{\linewidth}\raggedright
Trait
\end{minipage} & \begin{minipage}[b]{\linewidth}\raggedright
Construct
\end{minipage} & \begin{minipage}[b]{\linewidth}\raggedright
Illustrative Item
\end{minipage} \\
\midrule\noalign{}
\endhead
\bottomrule\noalign{}
\endlastfoot
Commitment to Work & Degree to which applicants prioritize long-term employment stability and are willing to exert effort beyond minimum job requirements & \emph{``I work harder than what my company requires of me.''} \\
Emotional Control & Ability to remain calm and patient in frustrating or interpersonally difficult workplace situations & \emph{``I remain calm even when someone at work does not treat me well.''} \\
Rule Adherence & Tendency to follow prescribed procedures fully and consistently, even under time pressure & \emph{``Even when I am in a hurry, I follow all the indicated steps in a procedure.''} \\
Ethics & Honesty and integrity in the workplace, including responsible use of company resources and truthfulness with supervisors & \emph{``I admit my mistakes at work even when doing so makes me look incompetent.''} \\
Collaboration & Prosocial orientation toward coworkers, including helpfulness and openness to people with different perspectives & \emph{``I always offer help to others without being asked.''} \\
Diligence & Persistence and personal initiative in achieving work goals, especially in the face of difficulty or added responsibility & \emph{``I cannot leave a task unfinished.''} \\
Safety at Work & Value placed on workplace safety practices, including participation in training and compliance with safety protocols & \emph{``Work safety is the responsibility of both the employer and the employees.''} \\
\end{longtable}

\section{Grouped SHAP Construction and Supplementary Profiles}\label{sec-grouped-shap-appendix}

\subsection{Construction}\label{construction}

Let \(\widehat q_f(x)\) denote firm model \(f\)'s predicted probability of attrition within 90 days. We treat the seven assessment domains as the players in the Shapley calculation. Each player contains one domain's composite score and its six constituent items, so those seven variables always enter or leave a synthetic prediction together.

To estimate the attribution for domain \(g\) for a focal application, each Monte Carlo draw first samples a predecessor-set size uniformly from \(\{0,\ldots,6\}\) and then samples a subset of that size uniformly from the other six domains. This reproduces the Shapley weighting obtained by drawing a uniformly random ordering and retaining the domains that precede \(g\). The draw also selects a background application with replacement from the 172,242 production applications. Starting from that background application's predictor vector, we replace the predecessor domains with the focal application's values and compute the change in predicted attrition when domain \(g\) is also replaced. Averaging these marginal contributions over 500 draws yields the grouped attribution.

All unrevealed domains in a draw come from the same background application, preserving their joint empirical distribution, while all variables within a revealed domain come from the focal application. The procedure does not estimate a conditional distribution for unrevealed domains given revealed ones. We use the same production population as the background distribution for all three models and split the 500 draws into ten deterministically seeded batches of 50.

Because the fitted models predict attrition, the computation first produces attrition-scale contributions. We multiply each contribution by \(-1\) to report retention-scale values. On that scale, the reference is each model's mean predicted retention in the production sample, and a positive value represents a positive contribution to the application's prediction relative to that reference. It does not describe the change that would result from increasing the domain score.

\subsection{Simulation Precision and Statistical Uncertainty}\label{simulation-precision-and-statistical-uncertainty}

We assess simulation precision by comparing the first 250 draws with the full set of 500. Across the 21 firm-domain cells, the application-level estimates from 250 and 500 draws have Spearman correlations of at least 0.993. Increasing from 250 to 500 draws changes any plotted trait-quintile mean by at most 0.007 percentage points and any domain's mean absolute importance by at most 0.019 percentage points. The finite-simulation adding-up residual---the difference between an application's prediction deviation and the sum of its seven estimated contributions---has a mean absolute value of 0.76--0.89 percentage points across firms and a 95th percentile of 1.90--2.21 points. The attrition-to-retention sign conversion is exact to machine precision.

The confidence intervals in the grouped-attribution figures use 500 bootstrap resamples of applications within each plotted cell. The fitted models and the Monte Carlo attribution estimates are held fixed. The intervals therefore describe sampling variation in the reported cell means; they do not incorporate uncertainty from re-estimating the random forests or rerunning the Shapley simulation.

\subsection{Unsigned Importance by Domain-Specific Quintile}\label{unsigned-importance-by-domain-specific-quintile}

\begin{figure}[htb]

\centering{

\pandocbounded{\includegraphics[keepaspectratio]{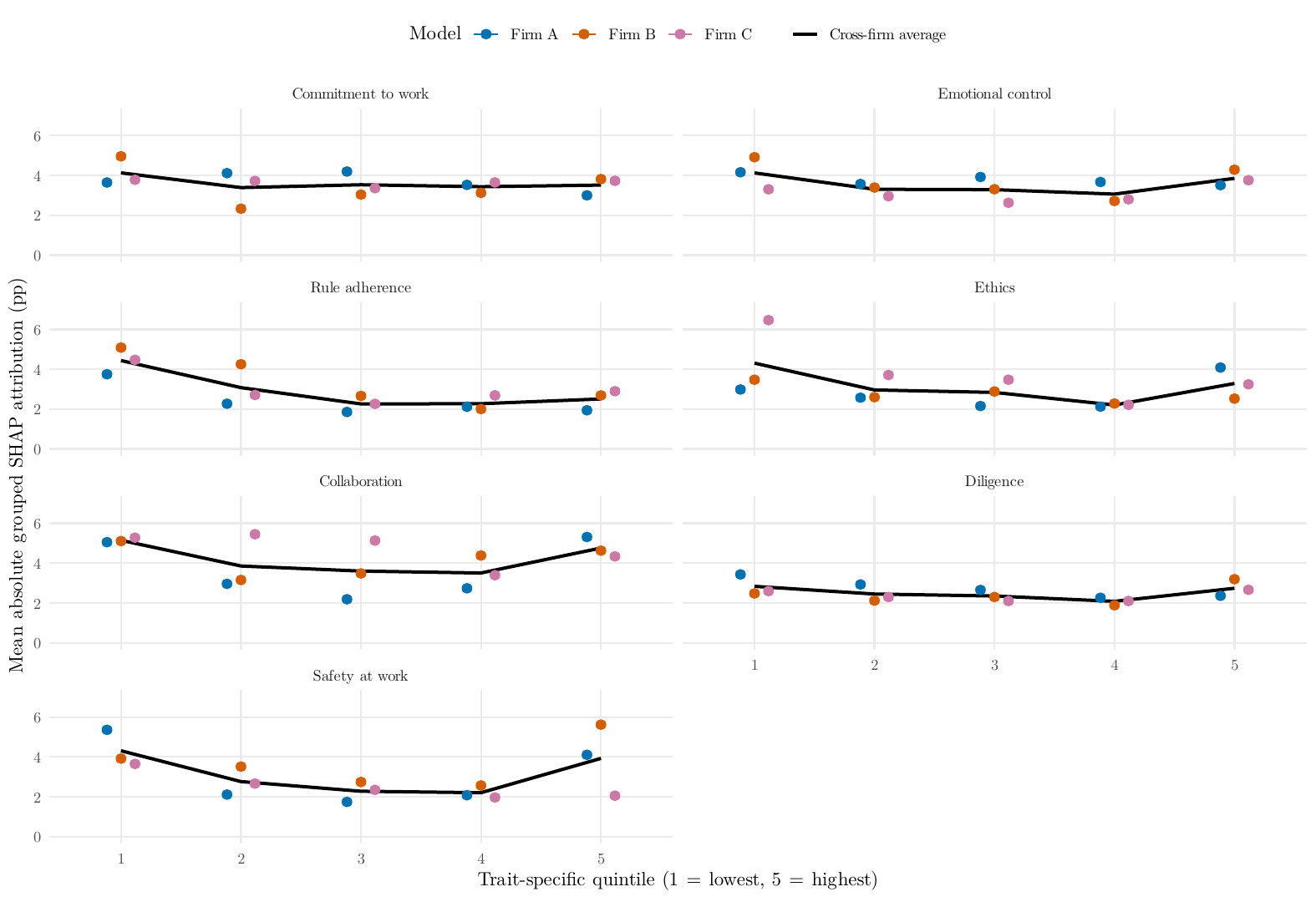}}

}

\caption{\label{fig-group-shap-importance-by-trait-quintile}Mean absolute grouped SHAP attributions by domain-specific score quintile and firm model in the production sample (\(N=172{,}242\)). Each panel corresponds to one assessment domain, with quintiles formed separately from that domain's composite score. Points report firm-specific mean absolute contributions to predicted 90-day retention relative to the corresponding model's mean prediction; error bars are 95\% bootstrap confidence intervals for the cell mean. The black line connects the unweighted mean across the three firm models within each domain and quintile. Attributions use 500 Monte Carlo draws per domain and application.}

\end{figure}

Figure~\ref{fig-group-shap-importance-by-trait-quintile} repeats the main-text comparison within quintiles of each domain's own score. Mean absolute attributions are generally larger at one or both tails of the score distribution than in its middle. The firm-specific markers nevertheless remain dispersed within quintiles, so the overall cross-firm averages do not eliminate heterogeneity in how strongly the models use a domain at a given score level.

\subsection{Unsigned Importance by Overall Assessment-Index Quintile}\label{unsigned-importance-by-overall-assessment-index-quintile}

\begin{figure}[htb]

\centering{

\pandocbounded{\includegraphics[keepaspectratio]{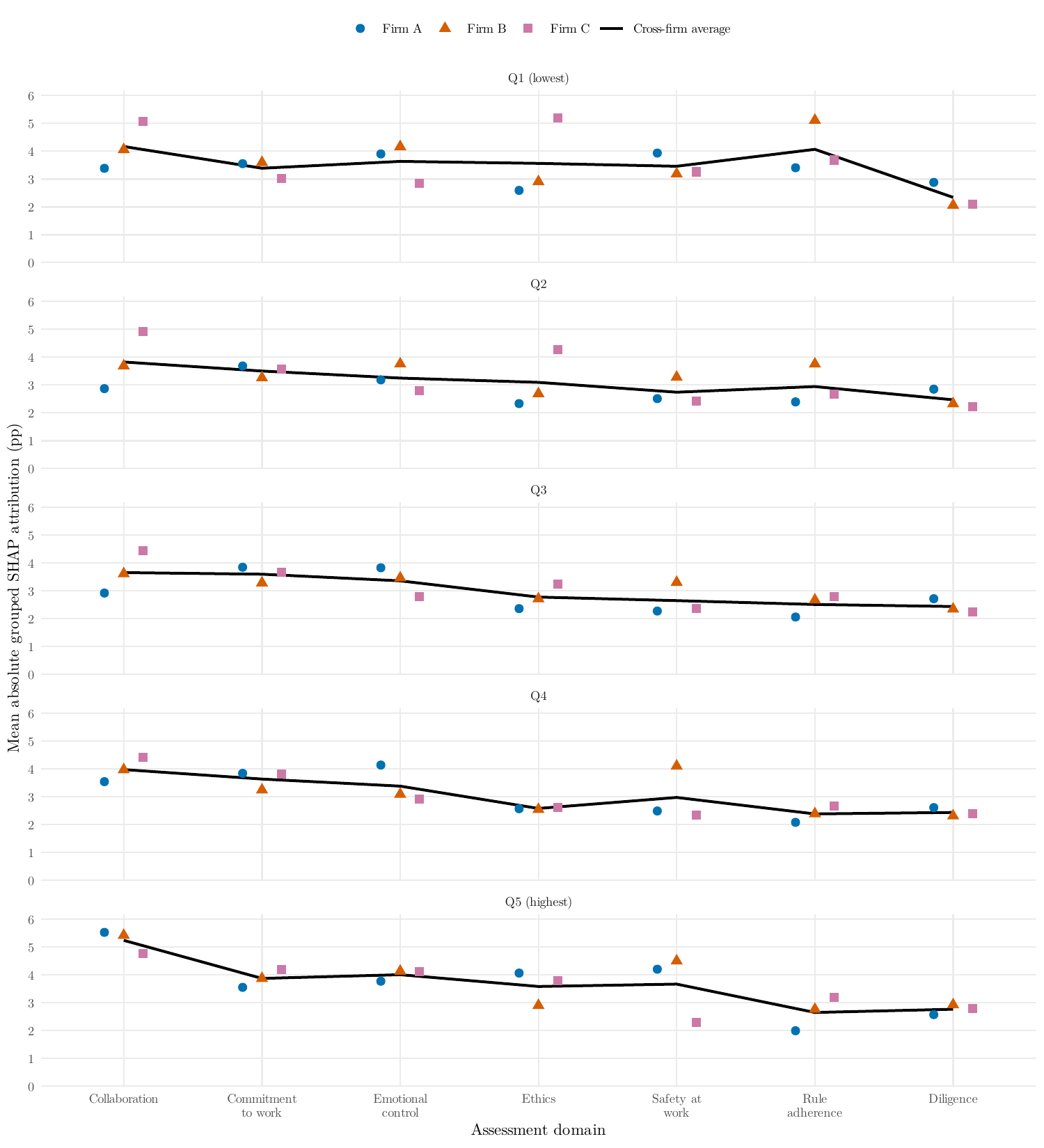}}

}

\caption{\label{fig-group-shap-importance-by-quality}Mean absolute grouped SHAP attributions by assessment domain, firm, and overall assessment-index quintile in the production sample (\(N=172{,}242\)). Each panel restricts the comparison to one assessment-index quintile. Markers report firm-specific mean absolute contributions to predicted 90-day retention relative to the model-specific mean prediction; the black line connects the unweighted mean across the three firm models within each domain and quintile. Domains retain the full-sample ordering used in the main text, and all panels use a common vertical scale. Attributions use 500 Monte Carlo draws per domain and application.}

\end{figure}

Figure~\ref{fig-group-shap-importance-by-quality} repeats the main-text importance comparison within quintiles of the overall assessment index. Across the second through fifth quintiles, the same three domains---collaboration, commitment to work, and emotional control---occupy the top three positions, with collaboration ranked first throughout. The bottom quintile differs: collaboration ranks first, followed by rule adherence and emotional control.

The firm-specific markers reveal heterogeneity that is obscured by these cross-firm averages. In every quintile, at least one domain's mean absolute attribution differs by more than 0.8 percentage points across firm models. Across quintiles, the largest within-domain firm spread ranges from 1.51 percentage points in the third quintile to 2.60 points in the first. The average ordering is thus similar over most of the assessment distribution even though the firm-specific functions continue to place different importance on the same domains within each quintile.

\subsection{Cross-Firm Agreement by Assessment-Index Quintile}\label{cross-firm-agreement-by-assessment-index-quintile}

\begin{figure}[htb]

\centering{

\pandocbounded{\includegraphics[keepaspectratio]{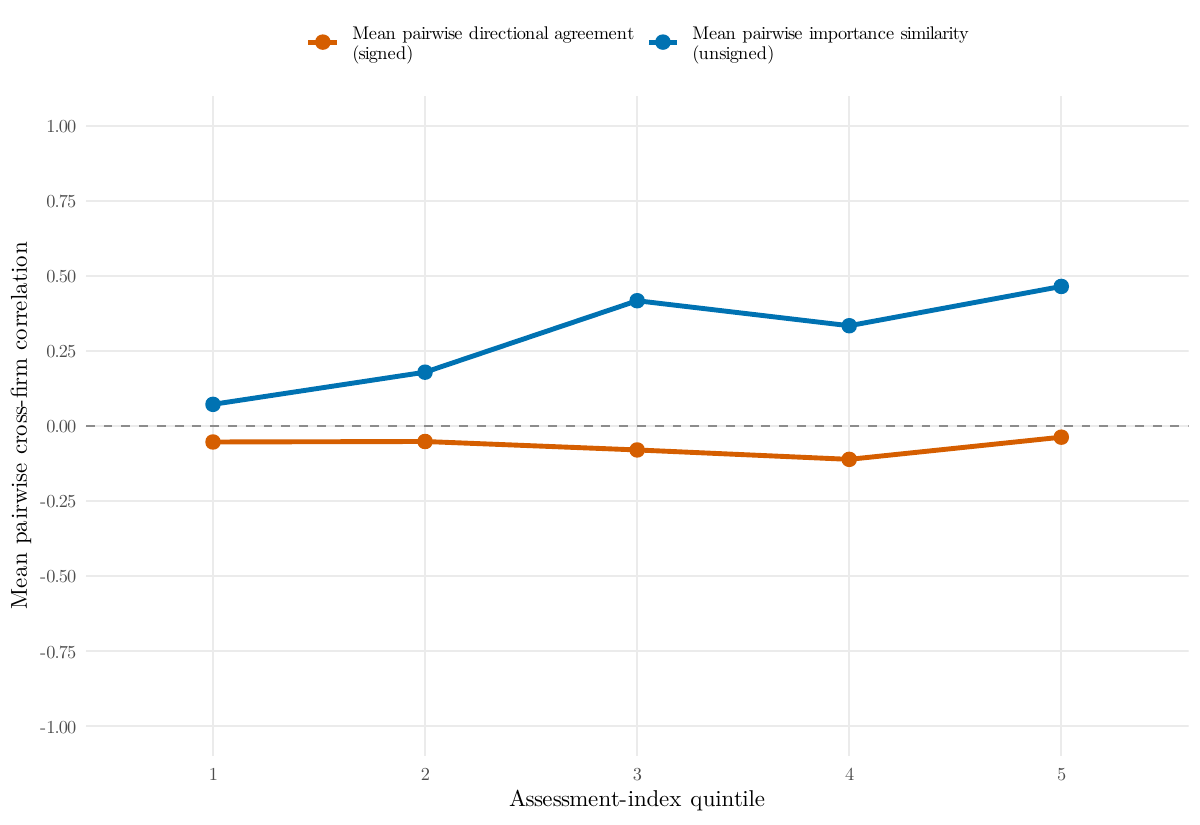}}

}

\caption{\label{fig-group-shap-agreement}Cross-firm agreement in grouped SHAP attributions by assessment-index quintile. Unsigned importance agreement is the rank correlation across the seven domains in mean absolute attribution. Signed agreement is the average within-domain rank correlation across applications. Each point averages the three firm pairs.}

\end{figure}

To summarize the importance differences in Figure~\ref{fig-group-shap-importance-by-quality} and compare them with disagreement in signed attributions, we compute two rank-correlation measures. For each assessment-index quintile and firm pair, unsigned importance agreement is the Spearman correlation across the seven domains in their mean absolute attributions. Signed agreement is computed within each domain as the Spearman correlation, across applications, between the two firms' signed attributions, and then averaged across domains. Figure~\ref{fig-group-shap-agreement} reports the mean of each statistic across the three firm pairs. Averaged across quintiles, unsigned importance agreement is 0.293 and signed agreement is \(-0.067\).

\subsection{Profiles by Overall Assessment-Index Quintile}\label{profiles-by-overall-assessment-index-quintile}

\begin{figure}[htb]

\centering{

\pandocbounded{\includegraphics[keepaspectratio]{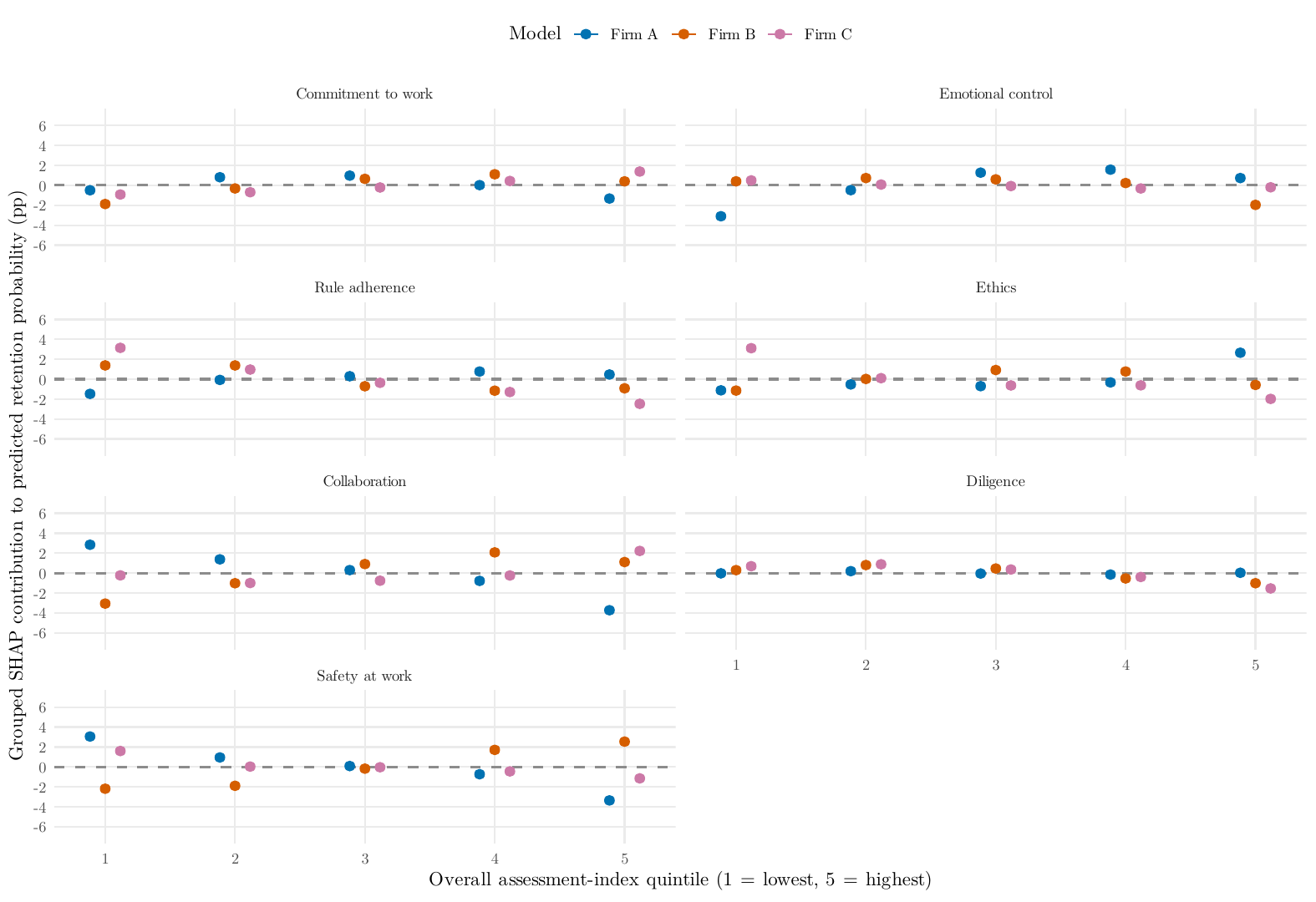}}

}

\caption{\label{fig-group-shap-overall-quality}Grouped SHAP profiles by overall assessment-index quintile in the production sample (\(N=172{,}242\); 500 Monte Carlo draws per domain and application). Each panel is an assessment domain. Points report firm-specific mean contributions to predicted retention relative to the model-specific reference prediction. Error bars are 95\% bootstrap confidence intervals for the cell mean, resampling applications within cells while holding the fitted models and attribution estimates fixed.}

\end{figure}

An alternative summary groups applications by the overall assessment index---the mean across the seven standardized domain scores---rather than by the level of a particular domain. Figure~\ref{fig-group-shap-overall-quality} reports the resulting profiles, with firm-specific estimates shown side-by-side and 95\% bootstrap confidence intervals.

\clearpage

\section{Domain-Score Contrasts}\label{sec-domain-target-details}

\subsection{Estimation and Inference}\label{estimand-and-inference-details}

\textbf{Samples and Target Population.}

The prediction models and score-contrast analysis use the same evaluation sample. It contains 8,119 applications: 1,682 at Firm A, 3,123 at Firm B, and 3,314 at Firm C.

Every evaluation application has an observed retention outcome and observed values for all 42 common assessment items and seven composite scores. These applications supply the assessment responses and retention outcomes used to estimate the firm-specific conditional retention relationships.

The production sample contains complete assessment responses, but no retention outcomes, for 172,242 applications. It therefore provides no information about the relationship between assessment responses and retention. Instead, it defines the common target population: production records determine which applications are eligible, how they are grouped, and how response configurations are averaged within each domain.

Formally, let \(P_f\) denote the evaluation distribution for firm \(f\), and let \(m_f(x)=E_{P_f}[Y\mid X=x]\). The distribution \(Q\) of assessment responses in the production population supplies the averaging distribution. Interpreting the resulting estimand as a retention association for production applications requires \(m_f(x)\) to continue to describe the conditional retention relationship in that population. The seven composite scores are deterministic sums of the 42 item responses and can therefore be recovered from \(X\).

\textbf{Estimator and Inference.}

The inferential target is the production-standardized finite contrast \(\Delta_{fgq}\), not an average derivative. We estimate it using automatic debiased machine learning, which provides inference for a researcher-specified functional of a flexibly estimated regression \citep{ChernozhukovNeweySingh2022}. For compact notation, let \(V_g=(S_g,Z_g)\), let \(T_gV_g=(S_g+0.25,Z_g)\), and define \(a_{gq}(V_g)=\mathbf{1}\{S_g\leq0.75,B_g=q\}\). Writing \(\pi_{gq}=E_Q[a_{gq}(V_g)]\), the main-text target can be written as

\[
\Delta_{fgq}
=
\frac{
E_Q\!\left[
a_{gq}(V_g)
\left\{
h_{fg}(T_gV_g)-h_{fg}(V_g)
\right\}
\right]
}{\pi_{gq}}.
\]

Within each outer fold, we Platt-calibrate the outcome regression using inner-fold out-of-sample predictions and evaluation outcomes before applying it to the held-out evaluation and production observations. Let \(b_{gq}(V_g)\) and \(\alpha_{fgq}(X)\) denote Riesz representers satisfying

\[
E_Q[b_{gq}(V_g)t(V_g)]
=
E_Q\!\left[
a_{gq}(V_g)
\{t(T_gV_g)-t(V_g)\}
\right],
\]

\[
E_{P_f}[\alpha_{fgq}(X)u(X)]
=
E_Q[b_{gq}(V_g)u(X)]
\]

for functions \(t\) and \(u\) in the fitted dictionaries. The cross-fitted estimator is

\[
\widehat{\Delta}_{fgq}
=
\frac{
\mathbb{P}_Q\!\left[
a_{gq}
\{\widehat h_{fg}(T_gV_g)-\widehat h_{fg}(V_g)\}
+
\widehat b_{gq}(V_g)
\{\widehat m_f(X)-\widehat h_{fg}(V_g)\}
\right]
+
\mathbb{P}_{P_f}\!\left[
\widehat\alpha_{fgq}(X)
\{Y-\widehat m_f(X)\}
\right]
}{
\mathbb{P}_Q[a_{gq}]
}.
\]

Here, \(\mathbb{P}_Q\) and \(\mathbb{P}_{P_f}\) denote held-out sample averages over the production and firm-specific evaluation samples, respectively. The evaluation contribution corrects estimation of the conditional retention relationship. The production contribution accounts for averaging over the target population, including uncertainty in the response distribution and in the share of applications in each eligible group.

We use four outer folds and reestimate every nuisance function within the corresponding training observations. The outcome regression uses the firm-specific random-forest settings and fold-specific Platt calibration described above. The conditional-averaging regression uses a 500-tree random forest with a minimum terminal-node size of 50 and square-root \texttt{mtry}, using the initial domain mean, the other 36 item responses, and the six other-domain composite scores. The two Riesz representers use fixed dictionaries of item indicators and pairwise products; the production representer additionally includes domain-score indicators and score-by-item interactions. Their penalty multiplier is selected from \(\{1.25,1,0.75,0.5\}\) by three-fold validation within each outer-training sample, with evaluation and production moments normalized separately.

Held-out evaluation and production score contributions are centered separately, and their covariance components are divided by the corresponding evaluation and production sample sizes. Because every firm comparison uses the same production applications, the covariance for firm differences retains the common production contribution. The simultaneous intervals reported in Appendix Figure~\ref{fig-domain-pairwise} use 1,999 Gaussian multiplier draws over the planned family of 105 pairwise comparisons.

The representer conditions require the shifted profiles entering \(T_gV_g\) to be supported under \(Q\) and the corresponding weighted production distribution to be represented in the support of \(P_f\). Inference treats the three firms as fixed and therefore does not characterize uncertainty over a broader population of firms. Asymptotic validity further requires sufficient support and appropriate approximation and convergence rates for the nuisance functions. The support diagnostic flags a target moment when a combination of dictionary regressors has effectively no variation in at least one fitting sample. The diagnostic isolates this support problem; validity continues to depend on the approximation and convergence conditions.

\textbf{Group Construction and Cutoff Stability.}

We construct the score groups after imposing the domain-specific eligibility restriction \(S_g\le0.75\). In the main specification, eligible applications are ordered by their initial domain score and assigned to five nonempty groups from lowest to highest. Identical scores remain together, so the groups approximate quintiles but can differ in size. Within a domain, all three firms use the same applications and group boundaries. Eligibility is determined separately for each domain, producing the sample sizes in Table~\ref{tbl-eligible-production}.

\begin{longtable}[]{@{}
  >{\raggedright\arraybackslash}p{(\linewidth - 4\tabcolsep) * \real{0.3333}}
  >{\raggedleft\arraybackslash}p{(\linewidth - 4\tabcolsep) * \real{0.3333}}
  >{\raggedleft\arraybackslash}p{(\linewidth - 4\tabcolsep) * \real{0.3333}}@{}}
\caption{Production applications eligible for a 0.25 increase in the domain mean. Eligibility requires an initial domain mean no greater than 0.75 and is defined separately by domain. Shares use all 172,242 production applications as the denominator. The counts describe the target population over which contrasts are averaged; retention outcomes come from the separate sample of 8,119 evaluation applications with complete item responses.}\label{tbl-eligible-production}\tabularnewline
\toprule\noalign{}
\begin{minipage}[b]{\linewidth}\raggedright
Trait
\end{minipage} & \begin{minipage}[b]{\linewidth}\raggedleft
Eligible production applications
\end{minipage} & \begin{minipage}[b]{\linewidth}\raggedleft
Share of production sample (\%)
\end{minipage} \\
\midrule\noalign{}
\endfirsthead
\toprule\noalign{}
\begin{minipage}[b]{\linewidth}\raggedright
Trait
\end{minipage} & \begin{minipage}[b]{\linewidth}\raggedleft
Eligible production applications
\end{minipage} & \begin{minipage}[b]{\linewidth}\raggedleft
Share of production sample (\%)
\end{minipage} \\
\midrule\noalign{}
\endhead
\bottomrule\noalign{}
\endlastfoot
Commitment to work & 131,126 & 76.1 \\
Emotional control & 106,925 & 62.1 \\
Rule adherence & 56,607 & 32.9 \\
Ethics & 68,169 & 39.6 \\
Collaboration & 106,975 & 62.1 \\
Diligence & 66,590 & 38.7 \\
Safety at work & 43,508 & 25.3 \\
\end{longtable}

The group cutoffs are selected from all eligible production applications without using retention outcomes. Inference treats these cutoffs as fixed and requires unique population cutoffs that remain unchanged under small changes in the score distribution. In 1,999 multinomial bootstrap draws, all four cutoffs remain unchanged in every draw for six domains and in 80.2 percent of draws for commitment. This exercise measures cutoff stability; it does not adjust the confidence intervals for cutoff uncertainty.

As an alternative, we group eligible applications by an overall raw assessment index rather than by the focal domain score. The index equals the mean of the seven raw domain means and, equivalently, the mean of all 42 item responses. It differs from the standardized composite-score index used in the earlier SHAP analysis. Although the two indices have a rank correlation of 0.999738 in the production sample, they can assign some applications to different groups. Because eligibility remains domain specific, the overall-index cutoffs can also differ across domains. Across these alternative groups, the share of bootstrap draws in which all four cutoffs remain unchanged ranges from 40.0 to 99.4 percent.

\subsection{Why Attributions and Score Contrasts Can Differ}\label{reading-an-attribution-and-a-contrast-for-the-same-group}

The 16,029 applications in the lowest eligible ethics group illustrate the distinction between the two empirical objects. Under Firm C's fitted model, their observed ethics responses contribute an average of +8.63 percentage points relative to the model's mean prediction over the full production sample. For the same applications, the estimated retention difference associated with a 0.25 increase in the ethics score is −6.64 percentage points, with a pointwise 95 percent confidence interval of {[}−8.52, −4.76{]}. The score-contrast estimate carries a support warning.

The two signs refer to different comparisons. The positive SHAP attribution says that the applications' observed ethics responses place their predictions above Firm C's reference prediction. The negative score contrast says that estimated average conditional retention is lower at a score 0.25 above these applications' initial ethics scores. The procedures also use different regression estimates: SHAP describes the fitted random forest, whereas the score contrast uses cross-fitting and bias correction to estimate the conditional retention relationship. The example therefore illustrates why the estimands can differ. It does not isolate that difference while holding the estimated regression fixed, and neither estimate validates the other.

\subsection{Pairwise Differences Across Firms}\label{pairwise-domain-contrasts}

\begin{figure}[H]

\centering{

\includegraphics[width=1\linewidth,height=\textheight,keepaspectratio]{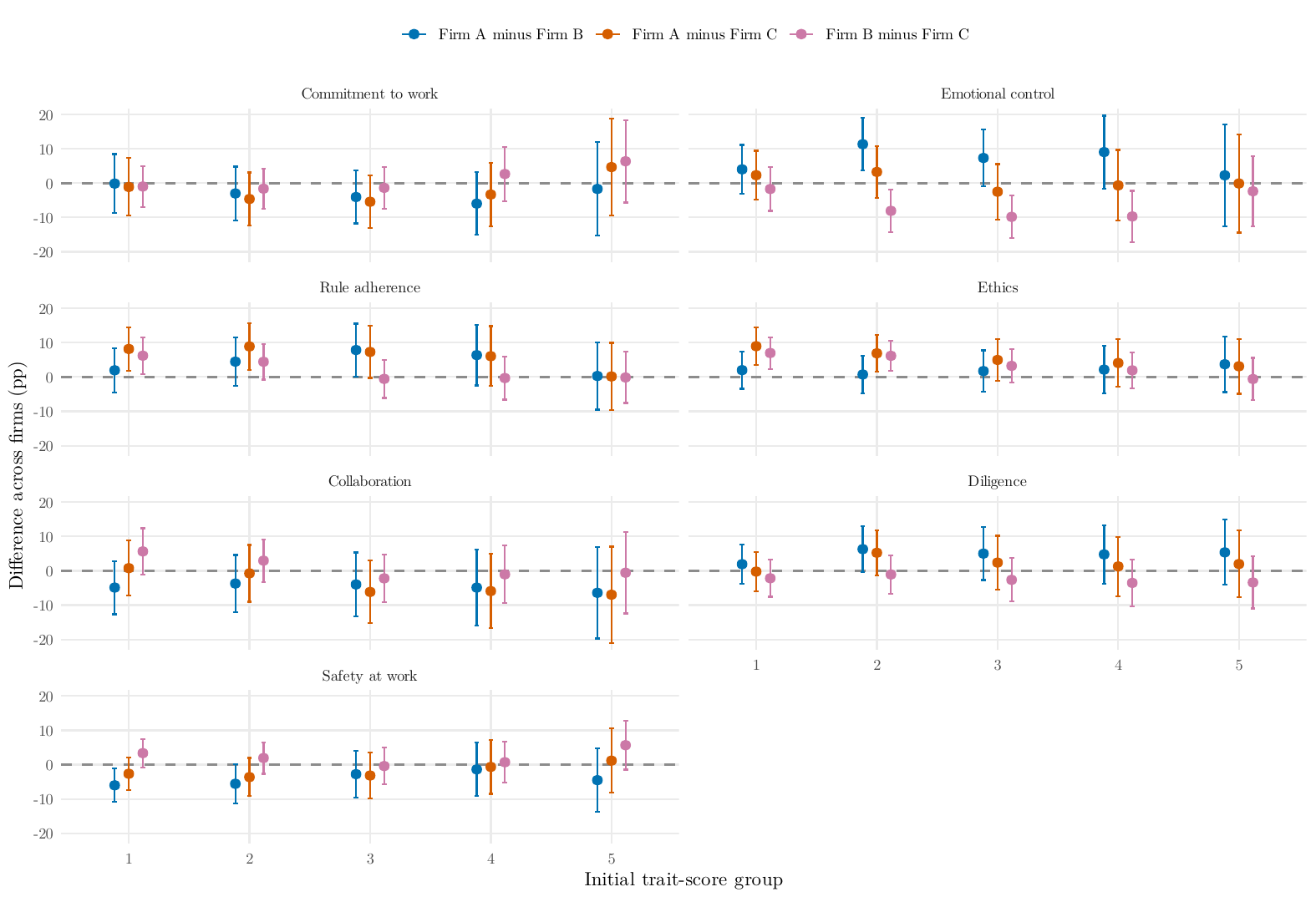}

}

\caption{\label{fig-domain-pairwise}Cross-firm differences in the retention association of a common domain-score increase, by assessment domain and initial domain-score group. Each point reports \(\Delta_{fgq}-\Delta_{hgq}\) in percentage points; positive values indicate that a 0.25 increase in the domain's mean recorded score has a more positive retention association at the first firm in the pair. Within each domain, eligible production applications have initial scores no greater than 0.75 and are divided into five groups from lowest to highest initial score. Bars are simultaneous 95 percent confidence intervals across all 105 comparisons; the dashed line marks zero. Evaluation outcomes estimate the firm-specific retention relationships, while production applications define the common target population and score groups. Seventy-six comparisons include at least one firm-specific estimate flagged by the support diagnostic, including nine of the thirteen whose intervals exclude zero.}

\end{figure}

Figure~\ref{fig-domain-pairwise} reports the pairwise differences \(\Delta_{fgq}-\Delta_{hgq}\) that underlie the cross-firm comparisons in the main text. Positive values indicate that the 0.25 score increase has a more positive retention association at the first firm in the pair. The bars are simultaneous confidence intervals across all 105 comparisons: seven domains, five initial-score groups, and three firm pairs. This figure, rather than overlap between the pointwise intervals in Figure~\ref{fig-domain-main}, provides the relevant test of cross-firm differences.

\subsection{Results by the Overall Assessment Index}\label{alternative-grouping-by-the-overall-raw-assessment-index}

\begin{figure}[H]

\centering{

\includegraphics[width=1\linewidth,height=\textheight,keepaspectratio]{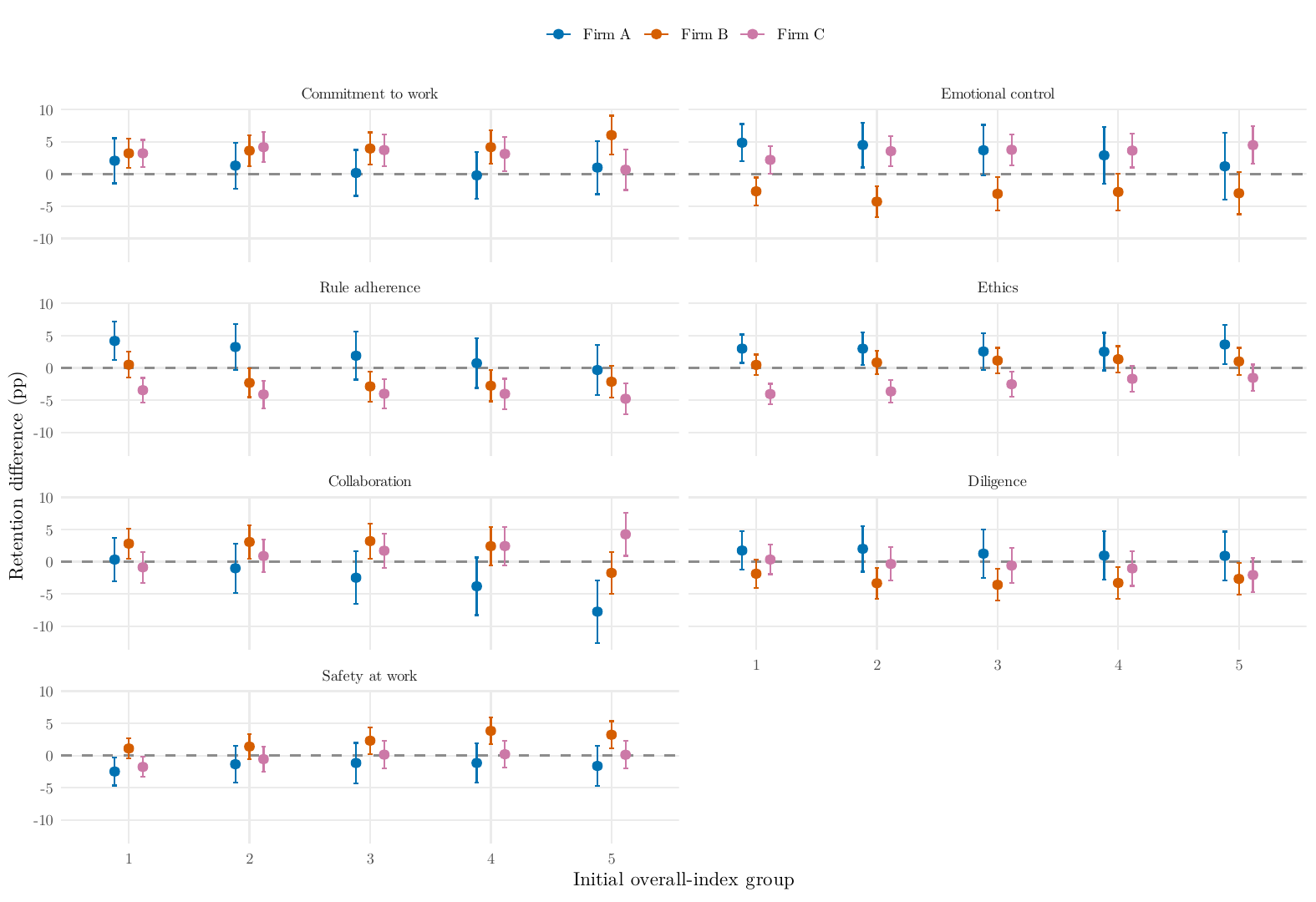}

}

\caption{\label{fig-overall-firms}Firm-specific retention associations for a common domain-score increase, by assessment domain and initial overall-assessment group. Each point reports \(\Delta_{fgq}\) in percentage points: the average retention difference associated with a 0.25 increase in the domain's mean recorded score, holding the other 36 responses fixed. Eligible production applications have initial domain scores no greater than 0.75 and are divided into five groups from lowest to highest overall assessment, defined as the mean of all 42 recorded responses. Bars are pointwise 95 percent confidence intervals; the dashed line marks zero. Evaluation outcomes estimate the firm-specific retention relationships, while production applications define the common target population, groups, and response distribution. Support diagnostics flag 47 of the 105 estimates.}

\end{figure}

Figure~\ref{fig-overall-firms} groups eligible applications by the overall raw assessment index rather than the focal domain score. Applications remain in their initial group throughout the comparison. The estimates therefore show how the same domain-specific score contrast varies across the broader assessment distribution; they do not represent movements between index groups. This exercise uses the same data as the main domain-score analysis and is not an independent replication. The figure reports all 105 firm-specific estimates, including those whose confidence intervals contain zero.

\end{document}

%% file: analysis/output/paper_evaluation/production/table1_restricted_samples.tex
\begin{table}[htbp]
\centering
\caption{Evaluation and Production Applications by Firm}
\label{tbl-table1}
\begin{tabular}{lrrrr}
\toprule
Statistic & Firm A & Firm B & Firm C & Total \\
\midrule
\multicolumn{5}{l}{\textit{Evaluation sample}} \\
Applications ($N$) & 1,682 & 3,123 & 3,314 & 8,119 \\
Retained beyond 90 days (\%) & 35.8 & 66.1 & 41.2 & 49.7 \\
Female (\%) & 19.6 & 54.3 & 58.8 & 48.9 \\
Age (years), mean & 33.3 & 25.3 & 24.4 & 26.6 \\
 & (8.4) & (4.8) & (4.1) & (6.5) \\
High school or above (\%) & 59.4 & 89.1 & 90.2 & 83.4 \\
\addlinespace
\multicolumn{5}{l}{\textit{Production sample}} \\
Applications ($N$) & 20,815 & 33,163 & 118,264 & 172,242 \\
Female (\%) & 23.9 & 59.2 & 63.5 & 57.8 \\
Age (years), mean & 35.2 & 25.6 & 25.9 & 27.0 \\
 & (9.0) & (6.0) & (5.9) & (7.1) \\
High school or above (\%) & 60.7 & 92.8 & 87.4 & 85.1 \\
\bottomrule
\end{tabular}
\begin{minipage}{0.96\linewidth}
\footnotesize\singlespacing
\textit{Notes:} The unit of observation is an application. Production applications have nonmissing responses to all 42 common assessment items and all seven composite scores. Evaluation applications require an observed binary 90-day attrition outcome, a hired and active rather than rejected status, no prior platform fitness score, and nonmissing responses to all 42 common assessment items and all seven composite scores. All statistics give equal weight to applications. Retention is reported only for evaluation applications and denotes tenure exceeding 90 days. Demographic statistics use all available observations and do not enter the predictive models. Nonmissing coverage across gender, age, and education ranges from 99.7 to 100.0 percent in the evaluation sample and from 94.3 to 100.0 percent in the production sample. Standard deviations for age are reported in parentheses. High school or above includes preparatoria or bachillerato, associate degrees, licenciatura or profesional degrees, and master's degrees.
\end{minipage}
\end{table}

%% file: analysis/output/paper_evaluation/production/figure_application_firm_matrix.tex
\begin{figure}[h]
\centering
\includegraphics[width=0.92\linewidth]{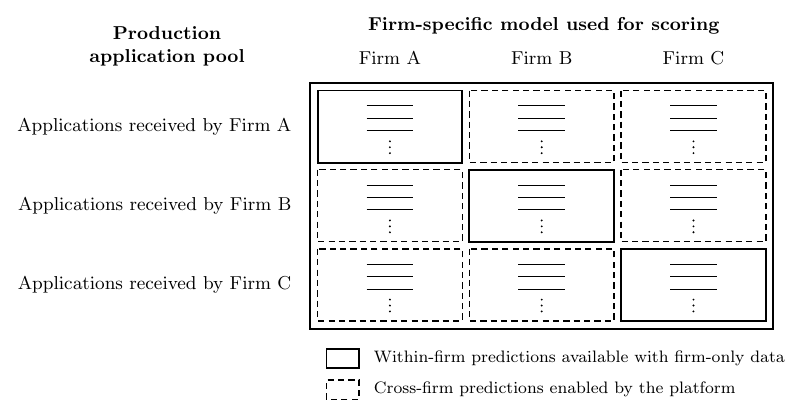}
\caption{The Application-by-Firm Prediction Matrix}
\label{fig-application-firm-matrix}
\begin{minipage}{0.94\linewidth}
\footnotesize\singlespacing
\textit{Notes:} Rows group production applications by the firm to which they were submitted; columns identify the firm-specific model used to generate predicted 90-day retention. Solid diagonal blocks contain the within-firm predictions that each firm can construct using its own data. Dashed off-diagonal blocks contain the cross-firm predictions made possible by the platform's common measurement architecture. Each block contains one predicted retention score for every application record in the corresponding application-pool--model pair. All entries are model-implied predictions, not observed retention outcomes under alternative firm assignments.
\end{minipage}
\end{figure}

%% file: analysis/output/paper_evaluation/candidate_model_comparison/table_candidate_model_auc.tex
\begin{table}[!htbp]
\centering
\caption{Cross-validated AUC by firm and candidate estimator}
\label{tab:candidate-model-auc}
\begin{tabular}{lccc}
\toprule
Estimator & Firm A & Firm B & Firm C \\
\midrule
Logistic Regression & 0.558 (0.002) & 0.534 (0.026) & 0.576 (0.007) \\
Random Forest & 0.858 (0.018) & 0.892 (0.005) & 0.916 (0.003) \\
Polynomial SVM & 0.819 (0.012) & 0.847 (0.010) & 0.877 (0.004) \\
XGBoost & 0.826 (0.013) & 0.844 (0.005) & 0.881 (0.006) \\
Neural Network & 0.826 (0.011) & 0.827 (0.004) & 0.855 (0.001) \\
\bottomrule
\end{tabular}
\begin{minipage}{0.95\linewidth}
\footnotesize Notes: Each cell reports mean 3-fold ROC AUC with its cross-fold standard error in parentheses. Tunable estimators use the highest-AUC restart within the firm-estimator pair. Random Forest, Polynomial SVM, XGBoost, and Neural Network use Bayesian optimization. These are candidate-comparison results used to choose the common estimator class; the separate nested assessment reports local predictive performance.
\end{minipage}
\end{table}

%% file: analysis/output/paper_evaluation/manuscript/tables/table_final_rf_hyperparameters.tex
\begin{table}[!htbp]
\centering
\caption{Final Random-Forest Hyperparameters by Firm}
\label{tab:firm-rf-hyperparameters}
\begin{tabular}{lrrrr}
\toprule
Model & Training applications & \texttt{mtry} & \texttt{min\_n} & Trees \\
\midrule
Firm A & 1,682 & 3 & 2 & 547 \\
Firm B & 3,123 & 16 & 1 & 2,469 \\
Firm C & 3,314 & 29 & 2 & 808 \\
\bottomrule
\end{tabular}
\begin{minipage}{0.95\linewidth}
\footnotesize Notes: Settings and training counts are copied from the audited final-model provenance. The manuscript uses only the three firm-specific models.
\end{minipage}
\end{table}

%% file: analysis/output/paper_evaluation/manuscript/tables/table_nested_cv_firm_models.tex
\begin{table}[!htbp]
\centering
\caption{Nested Cross-Validation Out-of-Fold AUC for Firm-Specific Random Forests}
\label{tab:paper-evaluation-nested-cv}
\begin{tabular}{lccccc}
\toprule
Model & OOF AUC & 95\% CI & Mean Fold AUC & SD Fold AUC & $N$ \\
\midrule
Firm A & 0.855 & [0.837, 0.873] & 0.855 & 0.031 & 1,682 \\
Firm B & 0.891 & [0.880, 0.902] & 0.891 & 0.007 & 3,123 \\
Firm C & 0.914 & [0.905, 0.923] & 0.915 & 0.005 & 3,314 \\
\bottomrule
\end{tabular}
\begin{minipage}{0.95\linewidth}
\footnotesize Notes: The outer loop provides held-out evaluation, while the inner loop tunes random-forest hyperparameters using Bayesian optimization within each outer-training split. OOF AUC and DeLong confidence intervals use the concatenated predictions from three held-out outer folds.
\end{minipage}
\end{table}

%% file: analysis/output/paper_evaluation/manuscript/tables/table_r2_cross_firm_auc.tex
\begin{table}
\centering
\begin{talltblr}[         
caption={Within- and Cross-Firm AUC of Firm-Specific Random Forest Models.\label{tbl-cross-firm-auc}},
note{}={Within-firm rows report the true nested-cross-validation out-of-fold AUC for each firm's model on its own evaluation sample. Cross-firm transfer rows report the AUC obtained by applying the final model trained for one firm to another firm's evaluation sample. Cross-firm AUC near 0.50 indicates chance-level prediction, confirming that the learned retention functions are firm-specific rather than portable across firms.},
]                     
{                     
colspec={Q[]Q[]Q[]Q[]Q[]Q[]},
hline{2}={1-6}{solid, black, 0.05em},
hline{1}={1-6}{solid, black, 0.08em},
hline{11}={1-6}{solid, black, 0.08em},
}                     
Training Firm & Evaluation Sample & Type & AUC & 95\% CI & N \\
Firm A & Firm A & Within-firm (nested CV) & 0.855 & [0.837, 0.873] & 1,682 \\
Firm A & Firm B & Cross-firm transfer & 0.507 & [0.485, 0.528] & 3,123 \\
Firm A & Firm C & Cross-firm transfer & 0.483 & [0.463, 0.502] & 3,314 \\
Firm B & Firm B & Within-firm (nested CV) & 0.891 & [0.880, 0.902] & 3,123 \\
Firm B & Firm A & Cross-firm transfer & 0.485 & [0.456, 0.512] & 1,682 \\
Firm B & Firm C & Cross-firm transfer & 0.473 & [0.453, 0.493] & 3,314 \\
Firm C & Firm C & Within-firm (nested CV) & 0.914 & [0.905, 0.923] & 3,314 \\
Firm C & Firm A & Cross-firm transfer & 0.494 & [0.466, 0.524] & 1,682 \\
Firm C & Firm B & Cross-firm transfer & 0.514 & [0.492, 0.536] & 3,123 \\
\end{talltblr}
\end{table}